\documentclass{aa}
\usepackage{graphicx}
\usepackage{txfonts}
\usepackage[normalem]{ulem}
\usepackage{color}
\usepackage{bm}
\usepackage{booktabs} % added by WHB: makes nicer looking tables
\usepackage[]{xcolor}
\usepackage{comment}
\usepackage{amsmath}
\begin{document}

   \title{
    Fundamental stellar parameters of benchmark stars \\
   from CHARA interferometry}
   \subtitle{II. Dwarf stars\thanks{Tables A.1.--A.9. are available at the CDS via anonymous ftp
to cdsarc.u-strasbg.fr (130.79.128.5)
or via http://cdsweb.u-strasbg.fr/cgi-bin/qcat?J/A+A/}} 
%   }}

 \author{I. Karovicova
          \inst{\ref{heidelberg}},
          %\and
                    T. R. White\inst{\ref{sydney},\ref{aarhus}},
                    T. Nordlander\inst{\ref{rsaa},\ref{astro3d}},
                    L. Casagrande\inst{\ref{rsaa},\ref{astro3d}},
                    M. Ireland\inst{\ref{rsaa}},
                    D. Huber\inst{\ref{honolulu}}
          }
 %I. Karovicova, T. R. White, T. Nordlander, L. Casagrande, M. Ireland, D. Huber
   \institute{Landessternwarte, University of Heidelberg
              K\"{o}nigstuhl 12, 69117, Heidelberg, Germany\\
              \email{karovicova@uni-heidelberg.de} \label{heidelberg}
                            \and 
        Sydney Institute for Astronomy (SIfA), School of Physics, University of Sydney, NSW 2006, Australia
         \label{sydney}
         \and
            Stellar Astrophysics Centre (SAC), Department of Physics and
            Astronomy, Aarhus University,
            Ny Munkegade 120, DK-8000 Aarhus~C, Denmark
        \label{aarhus}
        \and
                      Research School of Astronomy \& Astrophysics, Australian National University, Canberra, ACT 2611, Australia \label{rsaa}
                          \and
                      Center of Excellence for Astrophysics in Three Dimensions (ASTRO-3D), Australia \label{astro3d}
                                               \and  
          Institute for Astronomy, University of Hawai`i, 2680 Woodlawn Drive, Honolulu, HI 96822, USA \label{honolulu}
             }

   \date{Received July 20, 2021 ; accepted August 26 2021}

  \abstract{Stellar models applied to large stellar surveys of the Milky Way need to be properly tested against a sample of stars with highly reliable fundamental stellar parameters. We have established a programme aiming to deliver such a sample of stars.}
   {Here we present new fundamental stellar parameters of nine dwarf stars that will be used as benchmark stars for large stellar surveys. One of these stars is the solar-twin 18\,Sco, which is also one of the Gaia-ESO benchmarks. The goal is 
   %for all stars 
   to reach a precision of 1\% in effective temperature ($T_\mathrm{eff}$). This precision is important for accurate determinations of the full set of fundamental parameters and abundances of stars observed by the surveys.}
   {We observed  
   HD\,131156 ($\xi$\,Boo), HD\,146233 (18\,Sco), HD\,152391, HD\,173701, HD\,185395 ($\theta$\,Cyg), HD\,186408 (16\,Cyg\,A), HD\,186427 (16\,Cyg\,B), HD\,190360, and HD\,207978 (15\,Peg) using the high angular resolution optical interferometric instrument PAVO at the CHARA Array.
   We derived limb-darkening corrections from 3D model atmospheres and determined $T_\mathrm{eff}$ directly from the Stefan-Boltzmann relation, with an iterative procedure to interpolate over tables of bolometric corrections.
   Surface gravities were estimated from comparisons to Dartmouth stellar evolution model tracks. 
   We collected spectroscopic observations from the ELODIE spectrograph and estimated metallicities ($\mathrm{[Fe/H]}$) from a 1D non-local thermodynamic equilibrium (NLTE) abundance analysis of unblended lines of neutral and singly ionised iron.}
   {For eight of the nine stars we measure the $T_\mathrm{eff}$ $\lessapprox$\,1\%, and for one star better than 2\%. 
   We determined the median uncertainties in $\log\,g$ and $\mathrm{[Fe/H]}$
   %across our sample of stars are 
   as 0.015\,dex and 0.05\,dex, respectively.
    }
   {This study presents updated fundamental stellar parameters of nine dwarf stars that can be used as a new set of benchmarks. All the fundamental stellar parameters were based on consistently combining interferometric observations, 3D limb-darkening modelling, and spectroscopic analysis.
   %The stellar parameters in our study fulfil the ionization equilibria for all stars.
  The next paper 
  in this series 
  will extend our sample to giants in the metal-rich range.
   }

   
   \authorrunning{Karovicova et al.}
   \titlerunning{Fundamental stellar parameters of benchmark stars} 
   
   \maketitle
%
%-------------------------------------------------------------------

\section{Introduction}

Major efforts are underway to improve our understanding of stars, their populations, and thus the formation and evolution of our Galaxy. New instruments are enabling extensive surveys that allow us to explore the stellar content of the Milky Way in exquisite detail.

One of the latest and largest stellar surveys is being delivered by the Gaia satellite,
which is creating a precise 
3D map of more than a billion stars throughout our Galaxy \citep{gaia16}.
Complementary 
spectroscopic surveys such as APOGEE \citep{alendeprieto08}, GALAH \citep{desilva15}, Gaia-ESO \citep{Gilmore12,Randich13}, and the forthcoming 4MOST survey \citep{dejong12} are
providing the chemical compositions for a subset of Gaia stars. Overall, these large stellar surveys are helping us to improve
our knowledge of the chemo-dynamical evolution of our Galaxy.\\
Our understanding of stellar structure and evolution requires 
accurate measurements of fundamental stellar parameters such 
as the effective temperature ($T_\mathrm{eff}$), surface gravity 
($\log\,g$), metallicity ($\mathrm{[Fe/H]}$), and radius of the measured stars. 

In most cases fundamental stellar parameters are determined using
stellar spectroscopy. This unfortunately means that
the parameters are determined indirectly and are model dependent. 
This  negatively affects
the accuracy of delivered results and consistency between surveys.
Therefore, there is a great need for a set of reference stars  called benchmark stars 
for which these parameters can be measured across a wide range of 
parameter space in a direct and survey-independent way.\\

Optical interferometry offers a suitable solution because
it precisely measures the angular diameters of stars \citep{Boyajian12a, Boyajian12b, Boyajian13, vonbraun14, Ligi16, baines18, rabus19, rains20}.
The gold standard is then set by stars for which the effective temperature can be calculated directly 
from the measured angular diameter, $\theta$, and bolometric flux, $F_\mathrm{bol}$, 
according to the Stefan-Boltzmann law,
\begin{equation}
    T_\mathrm{eff} = \left(\frac{4F_\mathrm{bol}}{\sigma\theta^2} \right)^{1/4},
\end{equation}
where $\sigma$ is the Stefan-Boltzmann constant.
If reliable measurements of angular diameters can be achieved, then the quality of the derived $T_\mathrm{eff}$ will depend mainly on how well $F_\mathrm{bol}$ can be derived. 
 We note that $T_\mathrm{eff}$ is only weakly model-dependent via the adopted bolometric correction. For the stars analysed in this paper, a change of  2\% in bolometric fluxes impacts $T_\mathrm{eff}$ at the 30K level, which is a 0.5\% change.
Additionally, the 
linear radius can be directly determined from the angular diameter given the parallax.  
Optical interferometry thus allows for the determination of $T_\mathrm{eff}$, 
in a direct and also survey-independent way.
The stars with interferometrically measured angular diameters
can then be used to rigorously test and improve 
the models that are used when deriving stellar parameters from 
spectroscopic surveys. Additionally, the stars can be used as fundamental benchmark stars to calibrate the characterisation of exoplanet host stars \citep{tayar20}. The correct determination of fundamental 
parameters of benchmark stars is therefore extremely important.

The Gaia-ESO survey used a sample of 34 stars that had been interferometrically measured as benchmarks \citep{Jofre14,heiter15}.
Although the angular diameters of the stars in this sample
were observed using optical interferometry, and thus $T_\mathrm{eff}$ was determined directly, the observations were collected from the literature
from different instruments applying inconsistent limb-darkening corrections 
from various model atmosphere grids leading to an inhomogeneous dataset.
Therefore, there is an urgent need to refine the measurement of the current sample. Moreover, a wider parameter space needs to be further explored and thus the sample of benchmarks needs to be expanded. The current sample is rather small; a larger sample is needed to robustly test stellar models.

We have been working to improve and expand the sample of benchmark stars 
via a consistent approach. We measured angular diameters with the highest 
possible precision using a single state-of-the-art interferometric instrument, 
applying consistent limb-darkening corrections determined from the best available 
stellar models.  We resolved discrepancies between the spectroscopic, photometric, and interferometric $T_\mathrm{eff}$ of the metal-poor stars in the original Gaia-ESO benchmark sample \citep{karovicova18}. Recently, we presented the fundamental stellar parameters of a new and updated set of ten metal-poor benchmark stars \citep{karovicova20}. Here we present a second set of benchmarks from our sample, nine dwarf stars, where the fundamental stellar parameters of the stars were delivered applying the same methods as in the first sample.

        \begin{table*}%\small
  \caption{Stellar parameters}    
  \centering   
  \begin{tabular}{l l l l l l l r }      
\hline\hline   
Star     &  & Sp.type & Right ascension &    Declination &    m$_V$ &    m$_R$ &   $\pi$       \\ 
 & & & & & (mag) & (mag) & (mas)   \\
     \hline 
HD\,131156  & $\xi$ Boo A   & G7V &  14 51 23.3799 & $+$19 06 01.6994  & 4.59 & 3.91 & 148.520 $\pm$ 0.240 \\
HD\,146233  & 18 Sco       & G2V     &  16 15 37.2704 & $-$08 22 09.9820  & 5.50 & 5.50 & 70.768 $\pm$ 0.112  \\
HD\,152391  &              & G8.5V   &  16 52 58.8025 & $-$00 01 35.1163  & 6.64 & 6.66 & 59.538 $\pm$ 0.033  \\
HD\,173701  &              & G8V     &  18 44 35.1192 & $+$43 49 59.7891  & 7.52 & 7.52 & 36.978 $\pm$ 0.032  \\
HD\,185395  & $\theta$ Cyg & F3V    &  19 36 26.5343 & $+$50 13 15.9646  & 4.48 & 4.13 & 54.232 $\pm$ 0.186  \\
HD\,186408  & 16 Cyg A     & G1.5V   &  19 41 48.9539 & $+$50 31 30.2188  & 5.95 & 5.50 & 47.277 $\pm$ 0.033  \\
HD\,186427  & 16 Cyg B     & G3V     &  19 41 51.9732 & $+$50 31 03.0861  & 6.20 & 5.76 & 47.275 $\pm$ 0.025  \\
HD\,190360  &              & G7IV-V  &  20 03 37.4049 & $+$29 53 48.4953  & 5.71 & 5.20 & 62.444 $\pm$ 0.062  \\
HD\,207978  & 15 Peg       & F2V     &  21 52 29.9170 & $+$28 47 36.752   & 5.53 & 5.53 & 36.863 $\pm$ 0.097  \\ 
  \hline                                 
\end{tabular}
\tablefoot{Parallaxes are from Gaia DR2, without  zero point corrections. }
%\flushleft $^a$ The K5V companion is more than 5” away and makes a negligible contribution to the flux in PAVO, as evidenced by the the lack of a short baseline deficit in V$^2$ in Fig.\ref{figures1}.
\label{tab:parameters}
  \end{table*}

\begin{figure}
    \includegraphics[width=.5\textwidth]{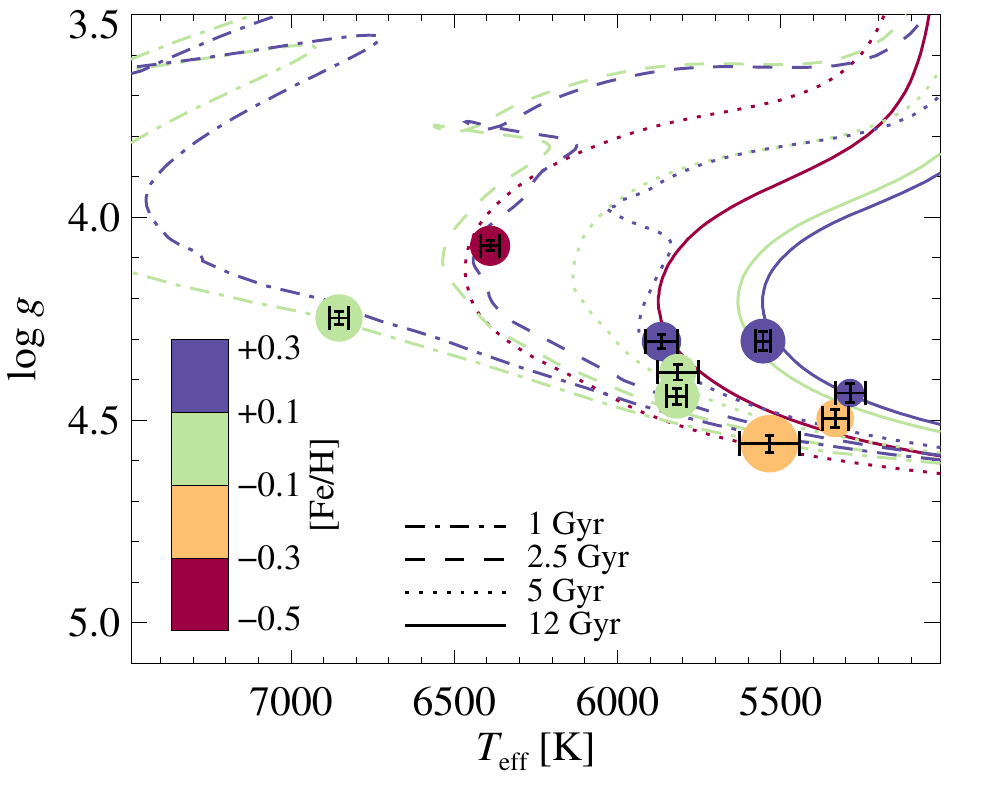}
    \caption{Stellar parameters of our programme stars, colour-coded by metallicity, compared to theoretical Dartmouth isochrones of different ages (line styles) and with metallicities $\rm$ $\mathrm{[Fe/H]}$ = $+0.2$, $0.0$ and $-0.5$ (colours). Formal 1$\sigma$ uncertainties are shown by the error bars. The symbol size is proportional to the angular diameter of each star.}
    \label{fig:HRdiag}
\end{figure}

      \begin{figure}
   \centering
       \includegraphics[width=\hsize]{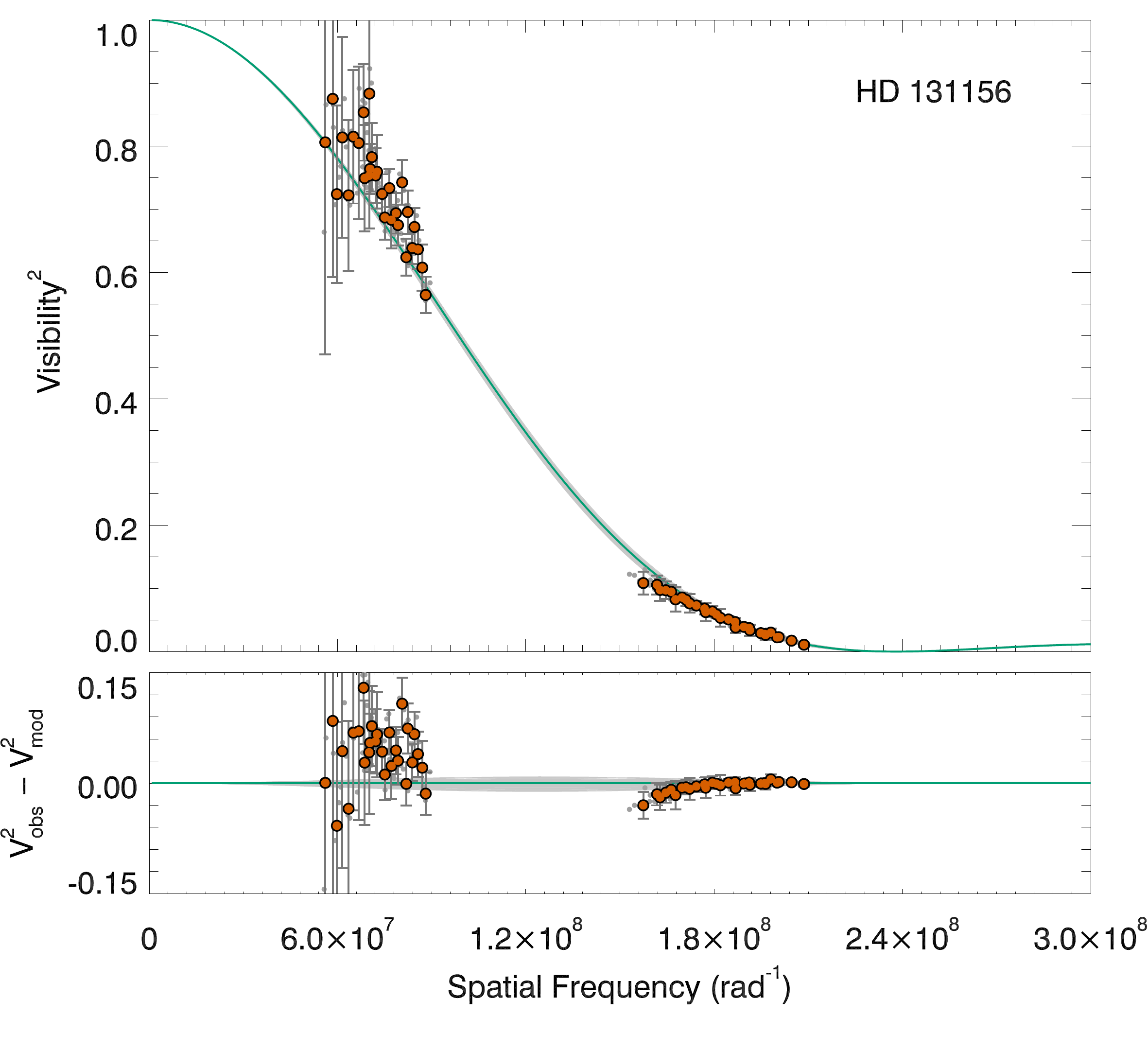}     
      \includegraphics[width=\hsize]{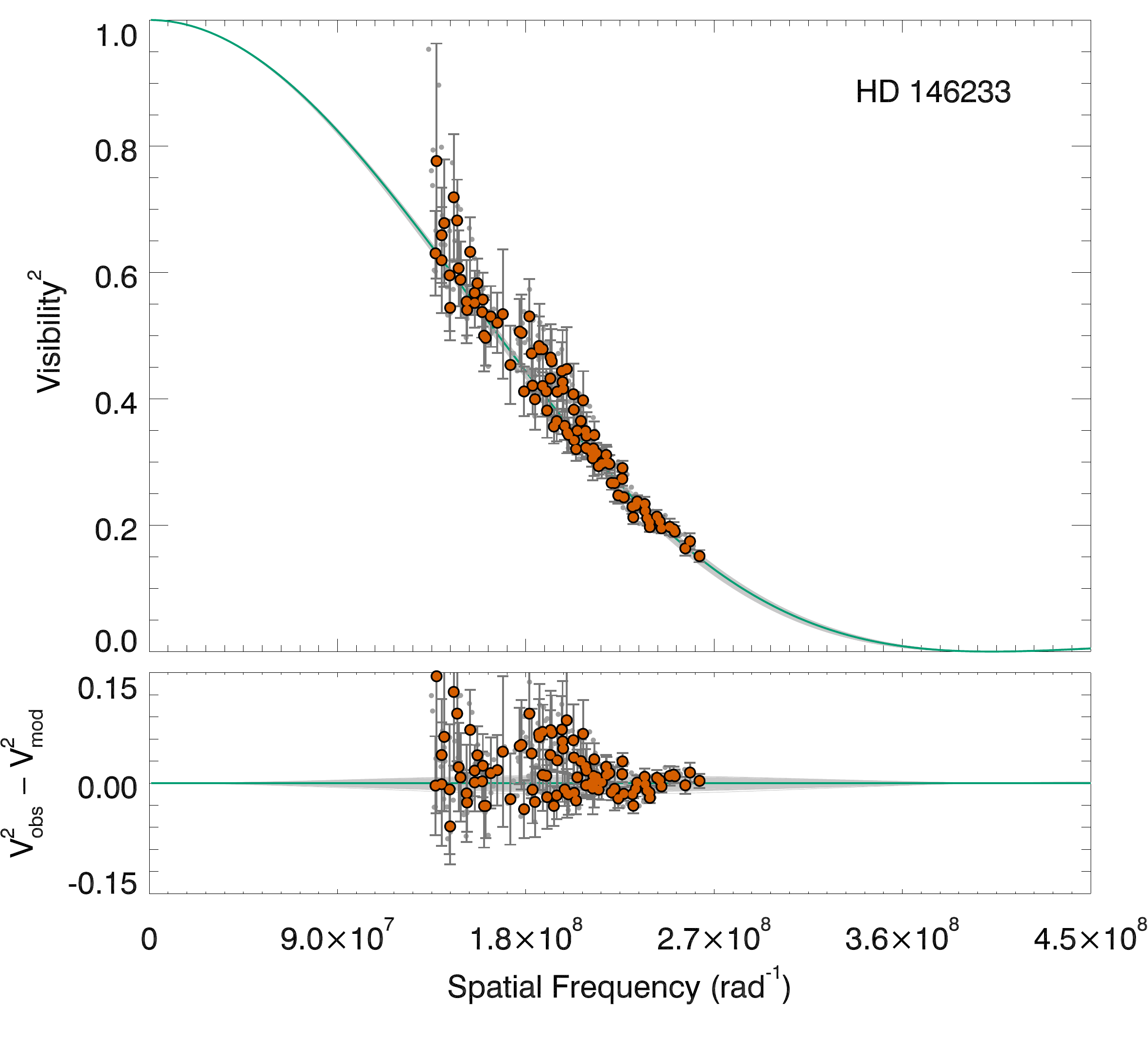}

      \caption{Squared visibility vs. spatial frequency for HD\,131156 and HD\,146233. The HD\, number is noted in the  upper right corner of each plot.
      The raw error bars were scaled to the reduced $\chi^{2}$ before the final fitting. For HD\,131156 the reduced $\chi^{2}$\,=\,4.7 
      and for HD\,146233 
      $\chi^{2}$\,=\,3.7.
      The grey dots are the individual PAVO measurements in each wavelength channel. 
      %The correlation between wavelength channels are taken into account by bootstrapping.
      For clarity, the weighted averages of the PAVO measurements are shown
      as red circles. The green line shows the fitted limb-darkened model to the PAVO data, with the light grey shaded region indicating the 1$\sigma$ uncertainties. The lower
      panel shows the residuals from the fit.
      }
      \label{figures1}
   \end{figure}

\section{Observations}

\subsection{Science targets}

We observed nine dwarf stars as a part of our programme which aims to deliver stars that can be used as benchmarks.
The dwarf stars will be added to the first set of metal-poor stars from this programme in \citet{karovicova20}. The parameters of the dwarf stars are listed in  Table~\ref{tab:parameters} and are shown in Fig~\ref{fig:HRdiag}.

Two stars (HD\,152391 and HD\,207978) were interferometrically observed for the first time. 
One of the stars (HD\,146233 (18\,Sco)) is a current Gaia-ESO benchmark star. The star was interferometrically observed by \citet{Bazot11} and \citet{Boyajian12a}. The Gaia-ESO benchmark sample adopted the angular diameter measured by \citet{Bazot11}; however, the measurement is in  stark disagreement with the measurement by \citet{Boyajian12a}, which is inconsistent with the status of solar twin. We are revising it here to confirm the angular diameter value of \citet{Bazot11}.
We discuss these previous measurements, and compare them with our new measurements in Sect.~\ref{comparison}.
Five of the other eight stars (HD\,131156, HD\,173701, HD\,185395, HD\,186408, and HD\,186427) have been proposed as future benchmarks by \citet{heiter15}. 

Three targets are in multiple systems (HD\,131156, HD\,186408, and HD\,186427). HD\,131156 ($\xi$\,Boo\,A) has a K5V companion which is more than 5\,arcsec away \citep{mason01} and makes a negligible contribution to the flux in Precision Astronomical
Visible Observations (PAVO) instrument, as evidenced by the  lack of a short baseline deficit in V$^2$ in Fig.\ref{figures1}. HD\,186408 (16\,Cyg\,A) is separated by approximately 40\,arcsec from HD\,186427 (16\,Cyg\,B), and both stars are observed as single objects by PAVO. HD\,186408 is itself a close binary, separated by approximately 3\,arcsec from a faint red dwarf companion \citep{raghavan06}. As for HD\,131156, this companion makes a negligible contribution to the flux. 

The metallicities of our nine dwarf stars range between $-0.5$ and $+0.2$\,dex. HD\,173701 and HD\,190360 both appear metal rich for their age in Fig.~\ref{fig:HRdiag}. We have integrated their kinematics over the past 3\,GYr using the MWPotential14 model within the galpy python package \citep{Bovy15} and find that the mean guiding radius of these stars to be 6.9 and 7.1\,kpc, respectively. This is broadly consistent with their higher metallicity and not atypical of outward migrations for $\sim$10-11\,Gyr  stars \citep{minchev18}.

\subsection{Interferometric observations and data reduction}\label{sec:observations}

Our interferometric observations were collected using the PAVO interferometric instrument \citep{Ireland2008} at the  Center for High Angular Resolution Astronomy (CHARA) array at Mt. Wilson Observatory, California \citep{tenBrummelaar05}.
PAVO is operating as a pupil-plane beam combiner in optical wavelengths
between $\sim$\,630--880\,nm. The limiting magnitude of the observed targets is 
R\,$\sim$\,7.5, which can be slightly extended up to R\,$\sim$\,8 if the weather conditions allow. 
The CHARA Array offers baselines up to 330\,m, thus making it
the longest available baselines in the optical wavelengths worldwide.

The stars were observed using baselines between 65.91\,m and 330.7\,m. 
We collected the observations between 
2013 July 6
and 2016 August 13. Additionally, several of our targets were previously   observed with PAVO between 2009--2012. We   included these observations in our study so that all data could be analysed consistently. In particular, our treatment of limb-darkening differs from the previous studies. The previously published data include one night of observations of 18\,Sco \citep{Bazot11}, which we supplement with three nights of new observations, as well as the observations of HD\,173701 \citep{huber12}, 16\,Cyg\,A and B, and $\theta$\,Cyg \citep{white13}.
Table~\ref{table:2} summarises the dates 
of all observations, telescope configurations, and the baselines, $B$ (distance between telescopes).
The data were reduced with the PAVO reduction software. 
The PAVO data reduction software has been used in multiple studies  \citep[e.g.][]{Bazot11,Derekas11,huber12,Maestro13} and it has been well tested, especially for single baseline squared visibility ($V^2$) observations. 

The data reduction software allows  the application of several visibility corrections. In particular, a coherence time ($t_0$) correction based on a ratio of coherent and incoherent visibility estimators may be used to correct visibility losses. The $t_0$ correction can introduce biases, and we have chosen not to use it. We note, however, that this correction was used in the study of 16 Cyg\,A and B, and $\theta$\,Cyg by \citet{white13}. Here we carry out a fresh data reduction for each target. We analyse the data consistently with the rest of the data in this study.

In comparison to the previous studies, we  changed the wavelength range we used, including all 38 channels, which range between 630 and 880\,nm. The previous studies only used the central 23 channels between 650 and 800\,nm because the channels at either end may be unreliable. However, after a careful investigation of our data, we found no such concerns, and so we included the full range in our analysis.

Immediately before and after the science targets a set of 
calibrating stars was observed
allowing us to monitor the interferometric transfer function.
These calibrating stars were selected from a catalogue  of CHARA calibrators 
and from the Hipparcos catalogue \citep{hipparcos}. 
Calibrators were selected to be unresolved or nearly unresolved sources that were
located close on sky to our targets. 
We determined the angular diameters of the calibrators 
using the $V-K$ relation of \citet{boyajian14}, and corrected for limb-darkening to determine the uniform disc diameter in the $R$ band.
We used $V$ band magnitudes from the Tycho-2 catalogue \citep{Hoeg2000} and converted into the Johnson system using the calibration by \citet{bessell00}. The $K$ band magnitudes were selected from the Two Micron All Sky Survey \citep[2MASS;][]{Skrutskie2006}. The reddening was estimated from the dust map of \citet{green15}, and the reddening law of \citet{odonnell94} was applied.
The relative uncertainty on calibrator diameters was set to 5\%
\citep{boyajian14}. This uncertainty covers both the uncertainty on the calibrator diameters as well as the reddening. The absolute uncertainty on the wavelength scale was set to 5\,nm.
 We checked all calibrators for
 binarity. 
 According to Gaia\,DR2, the proper motion anomaly \citep{kervella19}, the 
$\rm{\tt phot\_bp\_rp\_excess\_factor}$
 \citep{evans18}, and the renormalised unit weight error \citep[RUWE;][]{belokurov20} all suggest that none of our calibrators has a companion that
 is large enough to affect our interferometric measurements or estimated calibrator sizes.
For the calibrators, we used uniform disc angular sizes in the $R$ band.
We found that using 
high-order limb-darkening coefficients to estimate of the calibrator sizes 
has a negligible impact ($\sim$0.3\%), which is substantially smaller than our 5\% 
uncertainty in the calibrator diameters.
All the calibrating stars,
their spectral types, magnitudes in the $V$ and $R$ band, their expected angular diameters,
and the corresponding science targets are summarised in Table~\ref{table:3}.

%%%   missing 152391, 150727 3 scans,

 \begin{table*}[t!]
\caption{Interferometric observations -  Dwarfs}            
\label{table:2}      
\centering                          
\begin{tabular}{l l l c c l}      
\hline\hline            
Science target & UT date & Telescope & B (m) & No. of obs. & Calibrator stars\\  
\hline    
   HD\,131156 & 2015 Apr 28  & E1E2   & 65.91  & 2 & HD\,132145 \\
             & 2015 Apr 29  & E1E2   & 65.91  & 3 & HD\,132145, HD\,135263 \\                
             & 2015 May 2   & E2W2   & 156.26 & 3 & HD\,130878, HD\,135263 \\                 
   HD\,146233 & 2009 Jul 18  & S1W2   & 210.98 & 4 & HD\,145607, HD\,145788, HD\,147550 \\         
             & 2013 Jul 6   & E2W2   & 156.26 & 3 & HD\,145607, HD\,145788 \\ 
             & 2013 Jul 7   & W1W2   & 107.93 & 3 & HD\,145607, HD\,145788, HD\,147550 \\        
             & 2013 Jul 8   & E2W2   & 156.26 & 3 & HD\,145607, HD\,145788, HD\,147550 \\ 
   HD\,152391 & 2015 Jul 29  & S1E2   & 278.80 & 2 & HD\,149121, HD\,150379, HD\,154762 \\           
             & 2016 Aug 11  & E1W1   & 313.57 & 2 & HD\,154145, HD\,154445, HD\,154762 \\         
             & 2016 Aug 12  & E1W1   & 313.57 & 2 & HD\,151591, HD\,154145, HD\,154445 \\     
   HD\,173701 & 2010 Jul 19  & S2E2   & 248.13 & 3 & HD\,176131 \\   
             & 2011 Jul 2   & S1W1   & 278.50 & 2 & HD\,176131, HD\,176626, HD\,180681 \\     
             & 2012 Aug 6   & S1E2   & 278.80 & 1 & HD\,176131 \\     
             & 2012 Aug 8   & S1E1   & 330.70 & 3 & HD\,176131, HD\,180681 \\ 
   HD\,185395 & 2011 May 27  & E2W2   & 156.26 & 3 & HD\,188665, HD\,189296 \\ 
             & 2011 May 28  & E2W2   & 156.26 & 2 & HD\,177003 \\   
             & 2012 Aug 4   & S1W2   & 210.98 & 3 & HD\,176626, HD\,181960, HD\,184787, HD\,188665 \\   
             & 2012 Aug 6   & E2W2   & 156.26 & 4 & HD\,181960, HD\,184787, HD\,188665 \\  
             & 2012 Aug 8   & S1W2   & 210.98 & 3 & HD\,183142, HD\,184787, HD\,188665 \\   
             & 2012 Aug 10  & S2W2   & 177.45 & 3 & HD\,183142, HD\,184787, HD\,188665 \\   
             & 2012 Aug 11  & W1W2   & 107.93 & 1 & HD\,188665 \\      
   HD\,186408 & 2010 Jul 20  & S2E2   & 248.13 & 1 & HD\,179483 \\     
             & 2011 Jul 4   & S1W2   & 210.98 & 3 & HD\,177003, HD\,188252 \\   
             & 2011 Sep 9   & S2W2   & 177.45 & 3 & HD\,177003, HD\,181960 \\
             & 2012 Aug 4   & S1W2   & 210.98 & 3 & HD\,176626, HD\,181960, HD\,184787 \\   
             & 2012 Aug 8   & S1W2   & 210.98 & 3 & HD\,176626, HD\,183142, HD\,188665 \\  
             & 2012 Aug 9   & S2E2   & 248.13 & 3 & HD\,183142, HD\,184787 \\   
             & 2012 Aug 10  & S2W2   & 177.45 & 2 & HD\,177003, HD\,180681, HD\,188665 \\   
             & 2012 Aug 12  & E2W1   & 251.34 & 3 & HD\,183142, HD\,188665, HD\,190025 \\  
             & 2012 Aug 14  & S2E2   & 248.13 & 3 & HD\,183142, HD\,184787, HD\,188665 \\                
   HD\,186427 & 2011 Jul 4   & S1W2   & 210.98 & 3 & HD\,177003, HD\,188252 \\   
             & 2011 Sep 9   & S2W2   & 177.45 & 3 & HD\,177003, HD\,181960 \\     
             & 2012 Aug 4   & S1W2   & 210.98 & 3 & HD\,176626, HD\,181960, HD\,184787 \\   
             & 2012 Aug 8   & S1W2   & 210.98 & 3 & HD\,176626, HD\,183142, HD\,188665 \\  
             & 2012 Aug 9   & S2E2   & 248.13 & 3 & HD\,183142, HD\,184787 \\   
             & 2012 Aug 10  & S2W2   & 177.45 & 1 & HD\,177003, HD\,188665 \\   
             & 2012 Aug 12  & E2W1   & 251.34 & 3 & HD\,183142, HD\,188665, HD\,190025 \\  
             & 2012 Aug 14  & S2E2   & 248.13 & 3 & HD\,183142, HD\,184787, HD\,188665 \\  
   HD\,190360 & 2016 Aug 10  & E2W2   & 156.26 & 3 & HD\,191243, HD\,193553 \\      
             & 2016 Aug 13  & E1W2   & 221.85 & 2 & HD\,191243, HD\,193553 \\       
   HD\,207978 & 2015 Jul 27  & E2W2   & 156.26 & 2 & HD\,207071, HD\,208174 \\         
             & 2015 Jul 28  & E2W2   & 156.26 & 1 & HD\,207071, HD\,208174 \\          
             & 2015 Jul 28  & S1W2   & 210.98 & 3 & D 207071, HD\,207469, HD\,208174 \\    
             & 2015 Jul 29  & S2E2   & 248.13 & 5 & HD\,207071, HD\,213340 \\           
             & 2015 Sep 21  & E1W2   & 221.85 & 3 & HD\,207071, HD\,207469, HD\,213340 \\   
             & 2015 Sep 23  & E1W2   & 221.85 & 1 & HD\,209439, HD\,213340 \\            
             & 2015 Sep 25  & E2W2   & 156.26 & 3 & HD\,207469, HD\,208057, HD\,209439, \\  
\hline                                 
\end{tabular}
\end{table*}

      \begin{figure}
   \centering
   \includegraphics[width=\hsize]{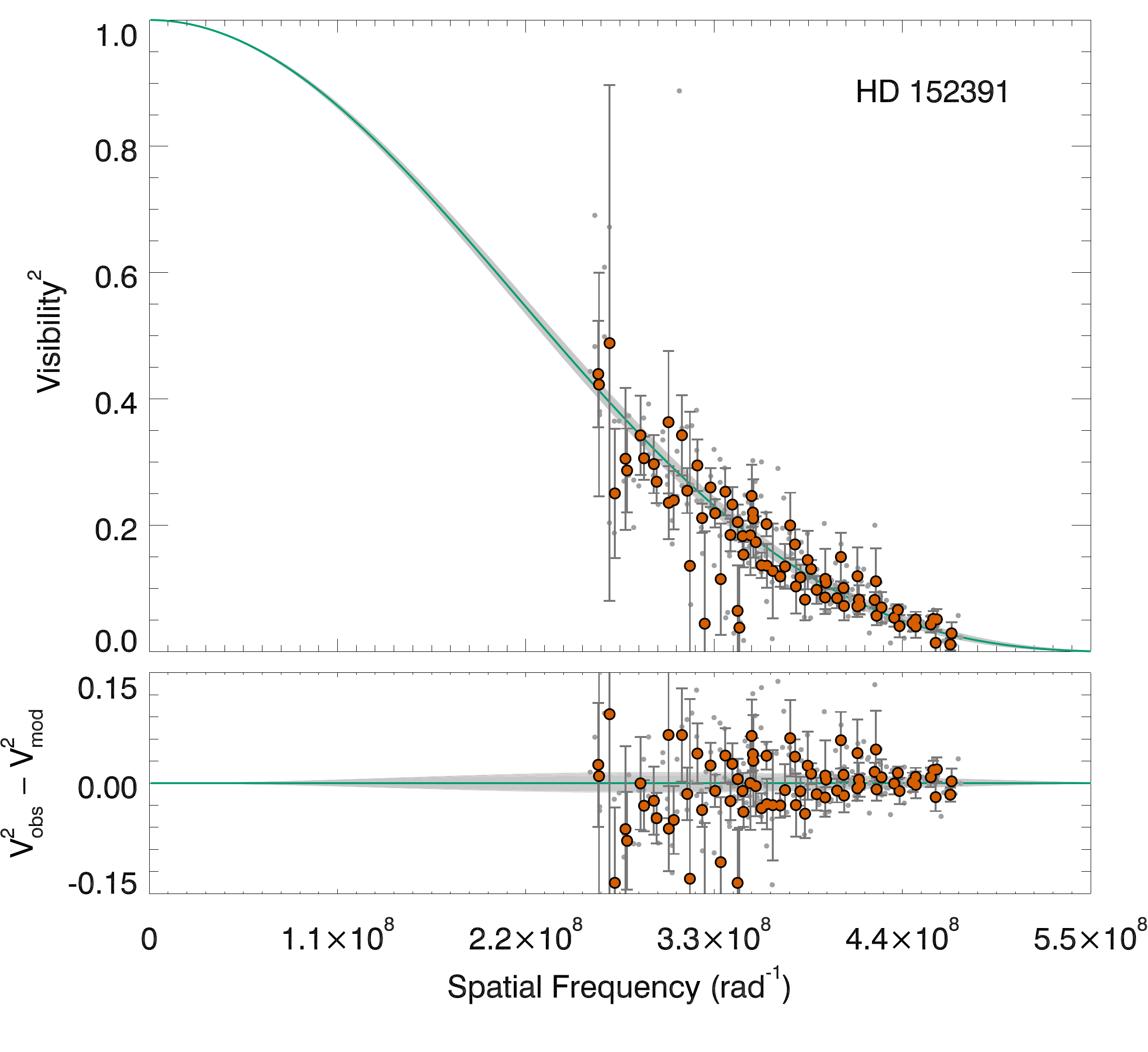} 
   \includegraphics[width=\hsize]{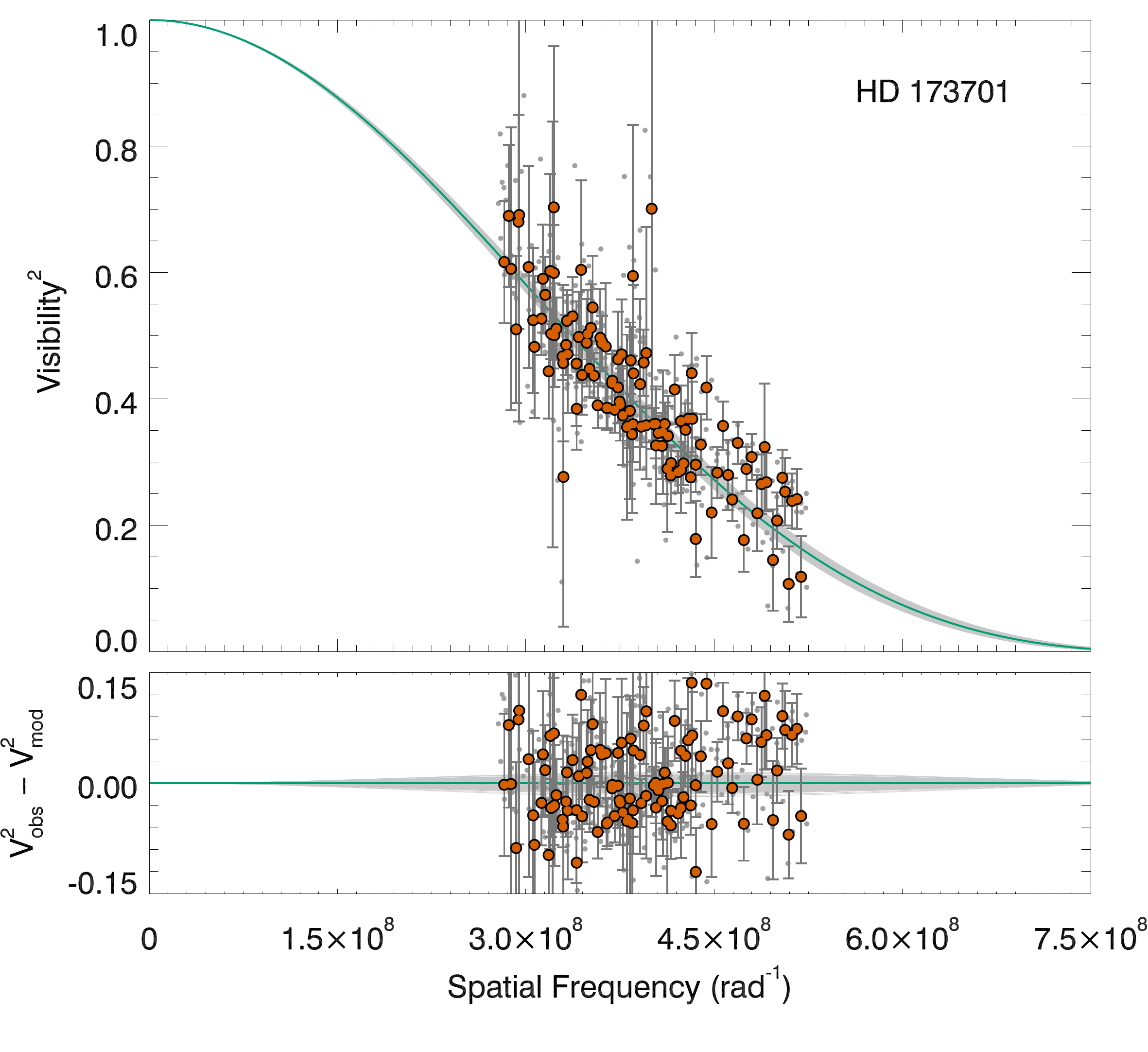}
    
      \caption{Squared visibility vs. spatial frequency for HD\,152391 and
HD\,173701. The lower panel shows the residuals from the fit.
The error bars are scaled to the reduced $\chi^{2}$. For HD\,152391
the reduced $\chi^{2}$\,=\,2.4 
and for HD\,173701
$\chi^{2}$\,=\,2.2. 
All lines and
symbols are the same as in Fig. \ref{figures1}.}
         \label{figures2}
   \end{figure}

      \begin{figure}
   \centering
      \includegraphics[width=\hsize]{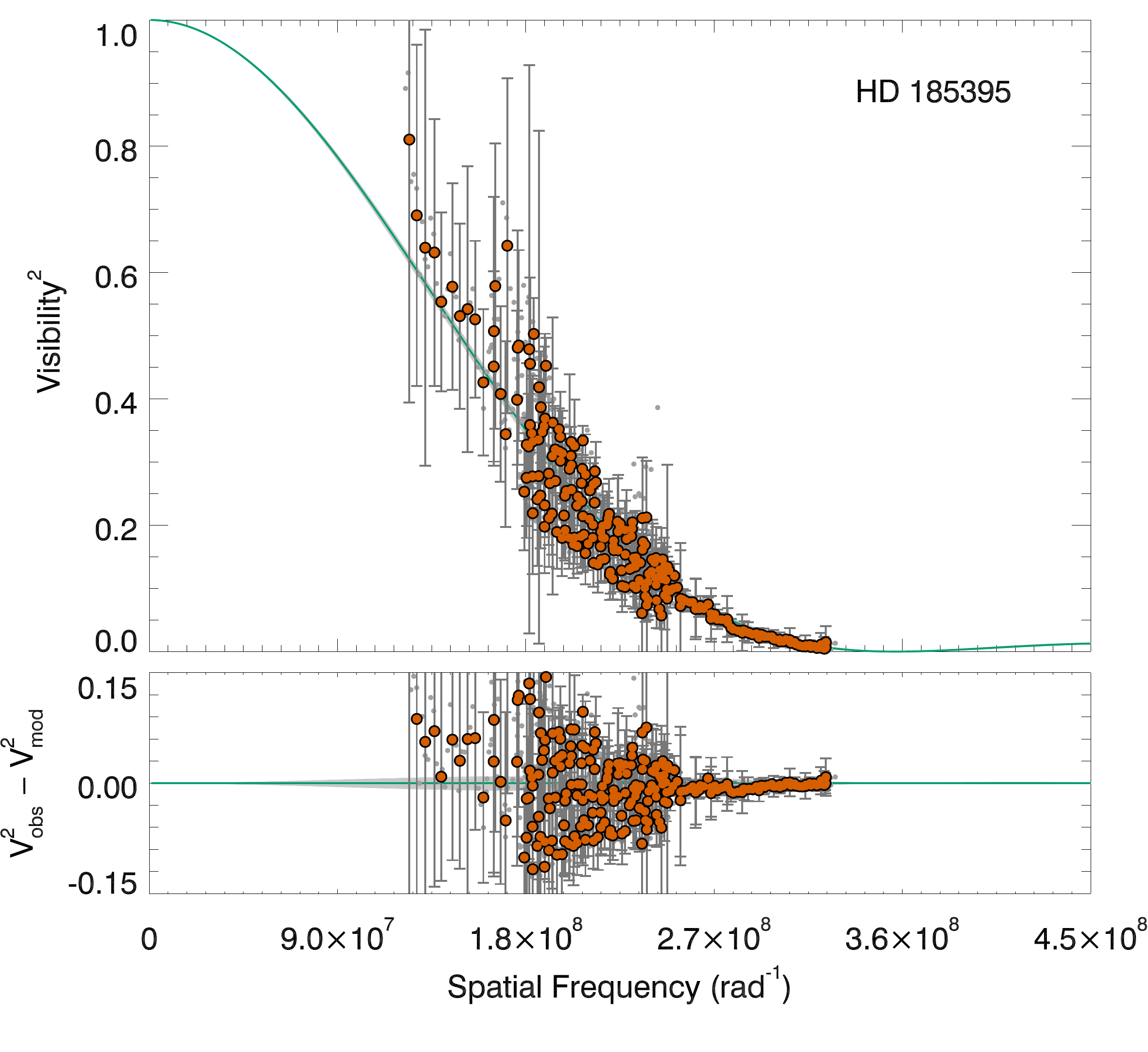}
   \includegraphics[width=\hsize]{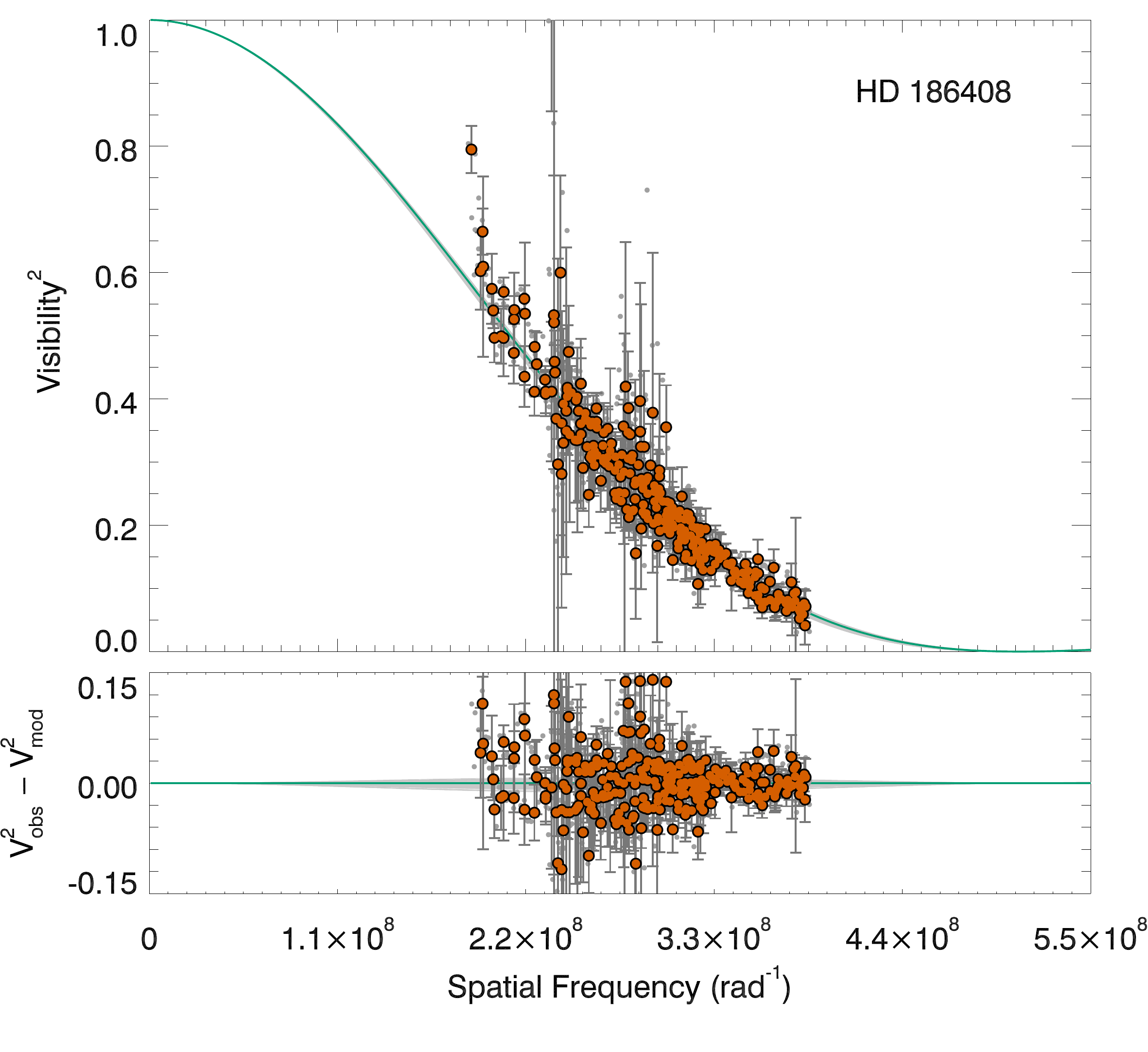} 

      \caption{Squared visibility vs. spatial frequency for HD\,185395 and
HD\,186408. The lower panel shows the residuals from the fit.
The error bars have been scaled to the reduced $\chi^{2}$. For HD\,185395
the reduced $\chi^{2}$\,=\,18.4 
and for HD\,186408
$\chi^{2}$\,=\,5.8. 
All lines and
symbols are the same as in Fig. \ref{figures1}. }
         \label{figures3}
   \end{figure}

    \begin{table}[t!]
\caption{Calibrator stars used for interferometric observations - dwarfs}            
\label{table:3}      
\centering                          
\begin{tabular}{l l c c c l}      
\hline\hline            
Calibrator & Spectral & m$_V$ & m$_K$ & $\mathrm{UD}$ \\ % & Science\\   
           & type         &       &       & (mas)         \\ % & target           \\
\hline
 HD\,130870  & B9      & 6.77 & 6.72 & 0.153 \\ % & HD\,131156 \\
 HD\,132145  & A1V     & 6.52 & 6.42 & 0.178 \\ % & HD\,131156 \\
 HD\,135263  & A2V     & 6.30 & 6.10 & 0.207 \\ % & HD\,131156 \\
 HD\,145607  & A2IV    & 5.42 & 5.05 & 0.327 \\ % & HD\,146233 \\   
 HD\,145788  & A0V     & 6.26 & 5.74 & 0.245 \\ % & HD\,146233 \\  
 HD\,147550  & B9V     & 6.24 & 5.96 & 0.215 \\ % & HD\,146233 \\   
 HD\,149121  & B9.5III & 5.63 & 5.68 & 0.240 \\ % & HD\,152391 \\  
 HD\,150379  & B9V     & 6.93 & 6.66 & 0.158 \\ % & HD\,152391 \\  
 HD\,151591  & A7IV/V  & 7.17 & 6.50 & 0.179 \\ % & HD\,152391 \\ 
 HD\,154145  & A3V     & 6.71 & 6.22 & 0.189 \\ % & HD\,152391 \\  
 HD\,154445  & B1V     & 5.61 & 5.29 & 0.276 \\ % & HD\,152391 \\  
 HD\,154762  & B9      & 7.28 & 7.08 & 0.129 \\ % & HD\,152391 \\   
 HD\,176131  & A2V     & 7.08 & 6.75 & 0.155 \\ % & HD\,173701 \\        
 HD\,176626  & A2V     & 6.85 & 6.77 & 0.147 \\ % & HD\,173701 \\ 
 %HD\,176626  & A2V     & 6.85 & 6.77 & 0.147 \\ % & HD\,185395 \\
 %HD\,176626  & A2V     & 6.85 & 6.77 & 0.147 \\ % & HD\,186408 \\
 %HD\,176626  & A2V     & 6.85 & 6.77 & 0.147 \\ % & HD\,186427 \\
 HD\,177003  & B2.5IV  & 5.38 & 5.90 & 0.204 \\ % & HD\,185395 \\
 %HD\,177003  & B2.5IV  & 5.38 & 5.90 & 0.204 \\ % & HD\,186408 \\  
 %HD\,177003  & B2.5IV  & 5.38 & 5.90 & 0.204 \\ % & HD\,186427 \\
 HD\,179483  & A2V     & 7.20 & 6.90 & 0.143 \\ % & HD\,186408 \\ 
 HD\,179733  & A0III   & 7.53 & 7.29 & 0.118 \\ % & HD\,173701 \\  
 HD\,180681  & A0V     & 7.48 & 7.39 & 0.111 \\ % & HD\,173701 \\  
 %HD\,180681  & A0V     & 7.48 & 7.39 & 0.111 \\ % & HD\,186427 \\  
 HD\,181960  & A1V     & 6.23 & 6.11 & 0.201 \\ % & HD\,185395 \\
 %HD\,181960  & A1V     & 6.23 & 6.11 & 0.201 \\ % & HD\,186427 \\
 %HD\,181960  & A1V     & 6.23 & 6.11 & 0.201 \\ % & HD\,186408 \\  
 HD\,183142  & B8V     & 7.07 & 7.53 & 0.096 \\ % & HD\,185395 \\
 %HD\,183142  & B8V     & 7.07 & 7.53 & 0.096 \\ % & HD\,186408 \\
 %HD\,183142  & B8V     & 7.07 & 7.53 & 0.096 \\ % & HD\,186427 \\
 HD\,184787  & A0V     & 6.68 & 6.65 & 0.155 \\ % & HD\,185395 \\
 %HD\,184787  & A0V     & 6.68 & 6.65 & 0.155 \\ % & HD\,186427 \\
 %HD\,184787  & A0V     & 6.68 & 6.65 & 0.155 \\ % & HD\,186408 \\ 
 HD\,188252  & B2III   & 5.90 & 6.36 & 0.165 \\ % & HD\,186408 \\   
 %HD\,188252  & B2III   & 5.90 & 6.36 & 0.165 \\ % & HD\,186427 \\
 HD\,188665  & B5V     & 5.13 & 5.52 & 0.248 \\ % & HD\,185395 \\  
 %HD\,188665  & B5V     & 5.13 & 5.52 & 0.248 \\ % & HD\,186408 \\ 
 %HD\,188665  & B5V     & 5.13 & 5.52 & 0.248 \\ % & HD\,186427 \\
 HD\,189296  & A4V     & 6.16 & 5.91 & 0.225 \\ % & HD\,185395 \\   
 %HD\,190025  & B5V     & 7.53 & 7.78 & 0.087 \\ % & HD\,186408 \\ 
 HD\,190025  & B5V     & 7.53 & 7.78 & 0.087 \\ % & HD\,186427 \\
 HD\,191243  & B6I     & 6.14 & 5.68 & 0.187 \\ % & HD\,190360 \\   
 HD\,193553  & B8      & 6.77 & 7.18 & 0.112 \\ % & HD\,190360 \\  
 HD\,207071  & B8    & 6.56 & 6.73 & 0.146 \\ % & HD\,207978\\ 
 HD\,207469  & A0    & 6.82 & 6.69 & 0.151 \\ % & HD\,207978\\  
 HD\,208057  & B3V   & 5.08 & 5.55 & 0.242 \\ % & HD\,207978\\    
 HD\,208174  & A2    & 6.78 & 6.14 & 0.213 \\ % & HD\,207978\\   
 HD\,209439  & A3    & 6.98 & 6.25 & 0.204 \\ % & HD\,207978\\   
 HD\,213340  & A0    & 6.58 & 6.36 & 0.182 \\ % & HD\,207978\\  
 \hline                                 
\end{tabular}
\end{table}

      \begin{figure}
   \centering
 
   \includegraphics[width=\hsize]{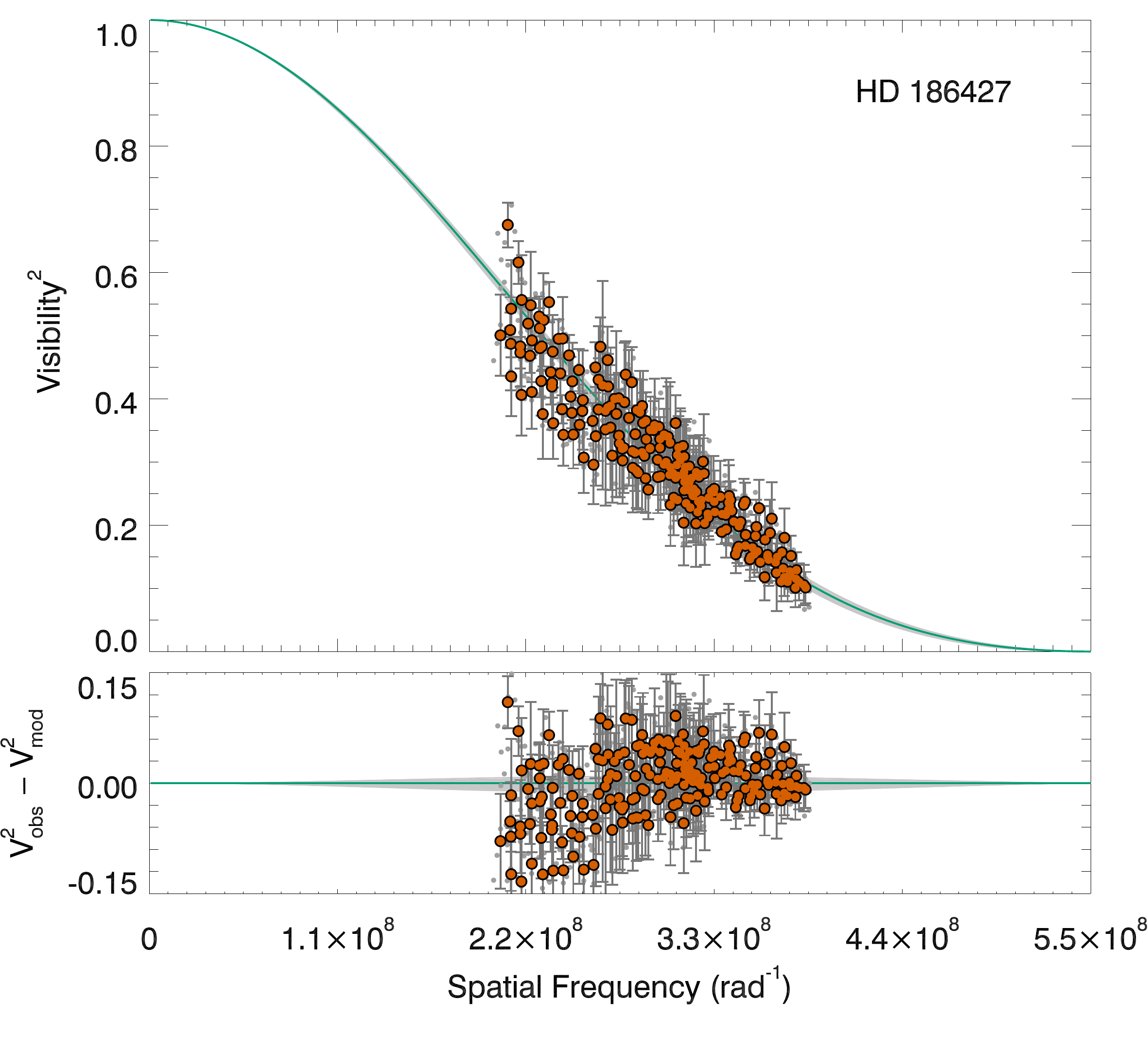}   
   \includegraphics[width=\hsize]{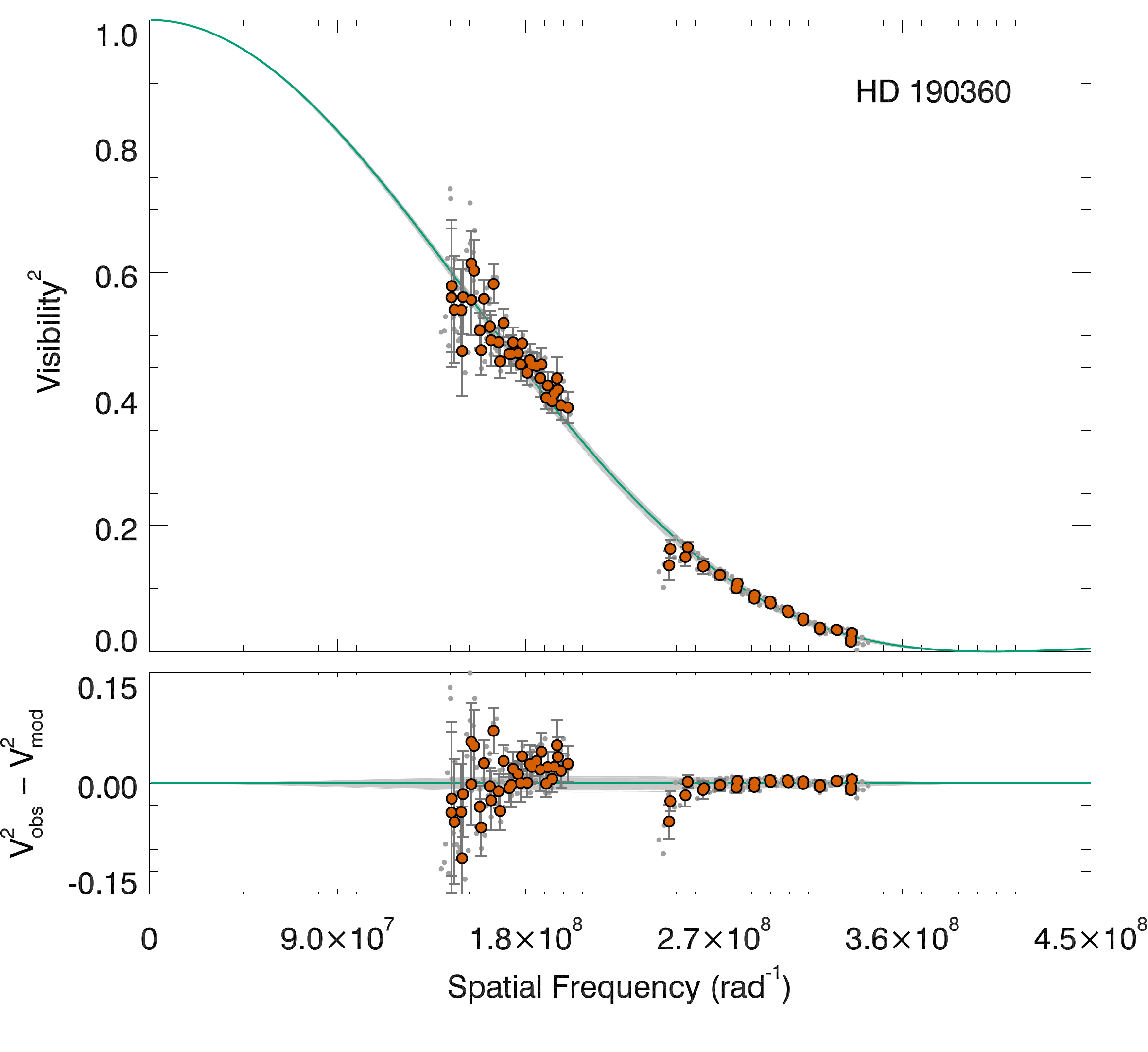}
      \caption{Squared visibility vs. spatial frequency for HD\,186427 and
HD\,190360. The lower panel shows the residuals from the fit.
The error bars are scaled to the reduced $\chi^{2}$. For HD\,186427
the reduced $\chi^{2}$\,=\,8.1 
and for HD\,190360
$\chi^{2}$\,=\,1.6. 
All lines and
symbols are the same as in Fig. \ref{figures1}.  }
         \label{figures4}
   \end{figure}

        \begin{figure}
   \centering
   \includegraphics[width=\hsize]{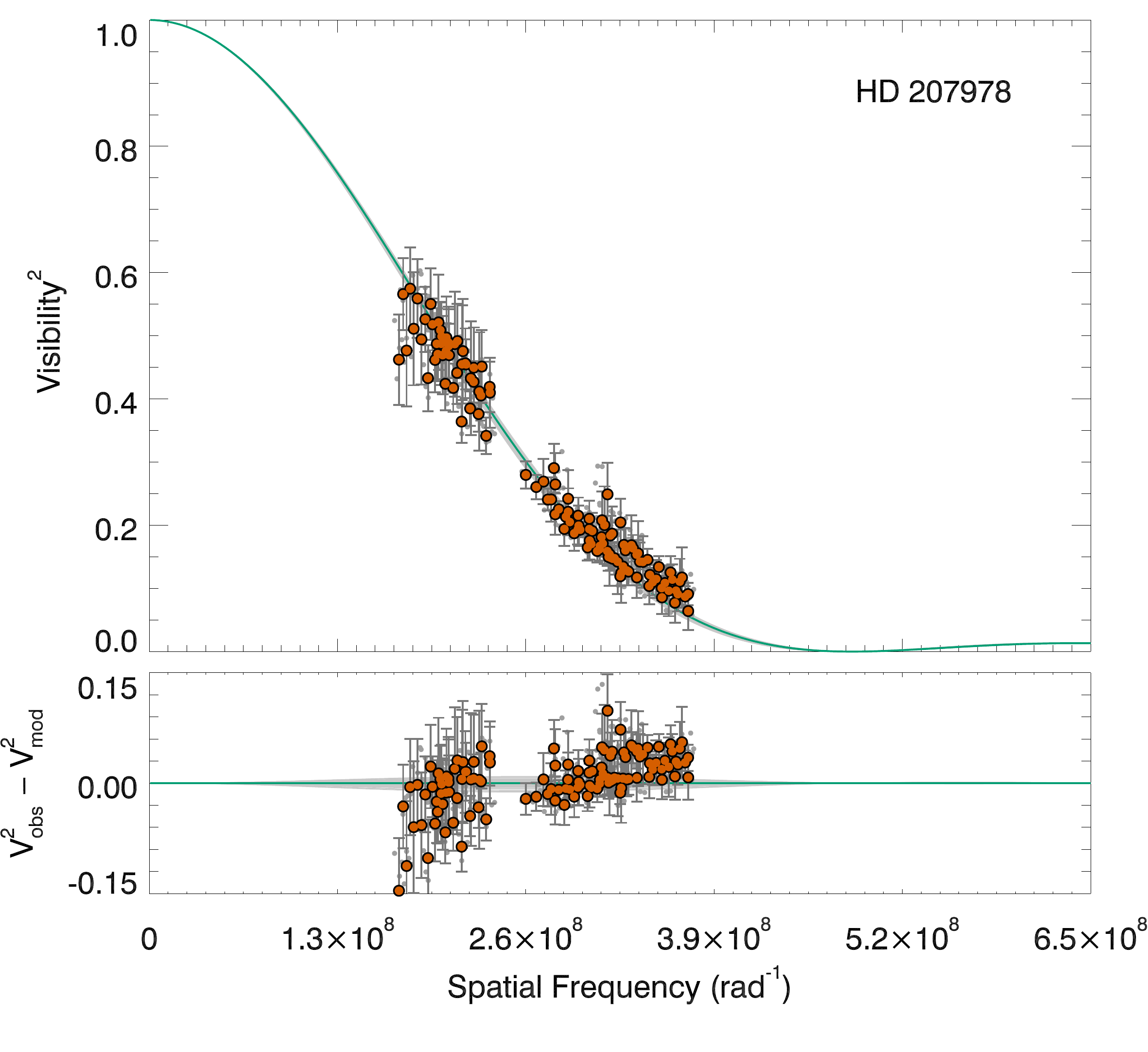} 
      \caption{Squared visibility vs. spatial frequency for HD\,207978. The lower panel shows the residuals from the fit.
The error bars are scaled to the reduced $\chi^{2}$. The reduced $\chi^{2}$\,=\,6.8. 
All lines and
symbols are the same as in Fig. \ref{figures1}.  }
         \label{figures5}
   \end{figure}

\begin{table}\small
\caption{Angular diameters and linear limb-darkening coefficients.}   \label{table:linear}      
\centering                          
\begin{tabular}{llll}      
\hline\hline            
Star  & $\theta_\mathrm{UD}$ (mas) & \multicolumn{2}{l}{Linear limb-darkening$^a$}   \\   
     &                              &    $u$    &    $\theta_\mathrm{LD}$ (mas)   \\
\hline    

HD\,131156  &   1.069 $\pm$ 0.008 & 0.596 $\pm$ 0.013 & 1.138 $\pm$ 0.009 \\
HD\,146233  &   0.633 $\pm$ 0.006 & 0.574 $\pm$ 0.013 & 0.669 $\pm$ 0.007 \\ 
HD\,152391  &   0.459 $\pm$ 0.006 & 0.616 $\pm$ 0.012 & 0.488 $\pm$ 0.007 \\
HD\,173701  &   0.313 $\pm$ 0.004 & 0.638 $\pm$ 0.012 & 0.333 $\pm$ 0.005 \\ 
HD\,185395  &   0.712 $\pm$ 0.006 & 0.489 $\pm$ 0.006 & 0.747 $\pm$ 0.006 \\
HD\,186408  &   0.503 $\pm$ 0.005 & 0.573 $\pm$ 0.013 & 0.531 $\pm$ 0.004 \\
HD\,186427  &   0.459 $\pm$ 0.005 & 0.577 $\pm$ 0.013 & 0.484 $\pm$ 0.005 \\
HD\,190360  &   0.631 $\pm$ 0.005 & 0.606 $\pm$ 0.012 & 0.671 $\pm$ 0.005 \\
HD\,207978  &   0.524 $\pm$ 0.005 & 0.501 $\pm$ 0.004 & 0.548 $\pm$ 0.006 \\

\hline                                 
\end{tabular}%\tablefoot{$^{(a)}$ Limb-darkening coefficients derived from the grid of \citet{magic15}; see text for details.}
\flushleft $^{a}$ Limb-darkening coefficients derived from the grid of \citet{claret11}; see text for details. The final limb-darkened diameters using the higher-order limb-darkening model are listed in Table 5.
\end{table}

      \begin{table*}%\small
   \caption{Observed ($\theta_{LD}$) and derived ($F_\mathrm{bol}$, $M$, $L$, $R$) stellar parameters.}    
   \centering   
  \begin{tabular}{l r r r r r}      
\hline\hline   
 Star  &     $\theta_{LD}$  &  $F_\mathrm{bol}$  &      $M$ (M$_\odot$)    &   $L$ (L$_\odot$) &    $R$ (R$_\odot$)\\
  &  (mas)  & (erg s$^{-1}$cm$^{-2}$10$^{-8}$) & &  \\
    \hline  
HD\,131156  & 1.124 $\pm$ 0.009  &  39.836 $\pm$ 2.568 & 0.88 $\pm$ 0.03 & 0.562 $\pm$ 0.036 & 0.817 $\pm$ 0.007 \\
HD\,146233  & 0.663 $\pm$ 0.007  &  16.711 $\pm$ 0.060 & 1.01 $\pm$ 0.02 & 1.043 $\pm$ 0.005 & 1.007 $\pm$ 0.011 \\
HD\,152391  & 0.477 $\pm$ 0.007  &  6.365 $\pm$ 0.105  & 0.83 $\pm$ 0.02 & 0.559 $\pm$ 0.009 & 0.878 $\pm$ 0.011 \\
HD\,173701  & 0.329 $\pm$ 0.005  &  2.836 $\pm$ 0.047  & 0.89 $\pm$ 0.02 & 0.646 $\pm$ 0.011 & 0.959 $\pm$ 0.015 \\
HD\,185395  & 0.737 $\pm$ 0.006  &  39.968 $\pm$ 0.187 & 1.40 $\pm$ 0.02 & 4.232 $\pm$ 0.035 & 1.462 $\pm$ 0.013 \\ 
HD\,186408  & 0.525 $\pm$ 0.004  &  10.845 $\pm$ 0.312 & 1.04 $\pm$ 0.03 & 1.511 $\pm$ 0.043 & 1.193 $\pm$ 0.009 \\
HD\,186427  & 0.479 $\pm$ 0.004  &  8.717 $\pm$ 0.325  & 1.02 $\pm$ 0.03 & 1.215 $\pm$ 0.045 & 1.088 $\pm$ 0.011 \\
HD\,190360  & 0.663 $\pm$ 0.005  &  13.947 $\pm$ 0.078 & 0.93 $\pm$ 0.02 & 1.114 $\pm$ 0.007 & 1.142 $\pm$ 0.009 \\
HD\,207978  & 0.542 $\pm$ 0.005  & 16.442 $\pm$ 0.067  & 1.07 $\pm$ 0.08 & 3.768 $\pm$ 0.025 & 1.587 $\pm$ 0.015 \\
  \hline                                 
\end{tabular}
\tablefoot{ $F_\mathrm{bol}$ and $L$ are obtained adopting L$_\odot=3.842 \times 10^{33}$ erg s$^{-1}$.}
\label{der_parameters}
  \end{table*}

{
\begin{sidewaystable*}
   \caption{Bolometric corrections}    
  \centering   
  \begin{tabular}{l l l l l l l l l l l l l l l l l l l}      
\hline\hline   
 & & & & & & & & & & & & & & & & & & \\
 Star        &     BC$_{H_p}$ &  BC$_{B_T}$ &  BC$_{V_T}$ &  BC$_J$ &  BC$_H$
      &  BC$_K$ &   $B_T$  &    e$B_T$    &  $V_T$   &  e$V_T$   &          $H_p$     & e$H_p$   &     $J$  &    e$J$  &    $H$  &    e$H$  &    $K$  &   e$K$ \\
 & & & & & & & & & & & & & & & & & & \\

 \hline

 & & & & & & & & & & & & & & & & & & \\
 HD\,131156   &     -0.234 & -0.985 & -0.179 & 1.183 &  1.515 & 1.607 &  5.575 & 0.014 & 4.757 & 0.009 &   4.6801 & 0.0014  &    -   &   -   &    -   &   -   &    -   &   -     \\
 & & & & & & & & & & & & & & & & & & \\
 HD\,146233   &     -0.172 & -0.832 & -0.114 & 1.095 &  1.369 & 1.454 &  6.292 & 0.015 & 5.570 & 0.009 &   5.6265 & 0.0005  &    -   &   -   &    -   &   -   &    -   &   -     \\
 & & & & & & & & & & & & & & & & & & \\
 HD\,152391   &     -0.286 & -1.135 & -0.232 & 1.254 &  1.626 & 1.728 &  7.594 & 0.015 & 6.733 & 0.010 &   6.7934 & 0.0026  &  5.235 & 0.035 &  4.942 & 0.044 &  4.835 & 0.029   \\
 & & & & & & & & & & & & & & & & & & \\
 HD\,173701   &     -0.291 & -1.201 & -0.237 & 1.277 &  1.645 & 1.754 &  8.606 & 0.016 & 7.610 & 0.011 &   7.6756 & 0.0013  &  6.088 & 0.021 &  5.751 & 0.016 &  5.670 & 0.021   \\
 & & & & & & & & & & & & & & & & & & \\
 HD\,185395   &     -0.068 & -0.425 & -0.020 & 0.742 &  0.873 & 0.936 &  4.942 & 0.014 & 4.534 & 0.009 &   4.5790 & 0.0004  &    -   &   -   &    -   &   -   &    -   &   -     \\
 & & & & & & & & & & & & & & & & & & \\
 HD\,186408   &     -0.160 & -0.813 & -0.101 & 1.084 &  1.346 & 1.432 &  6.719 & 0.014 & 5.997 & 0.009 &   6.1084 & 0.0007  &    -   &   -   &    -   &   -   &  4.426 & 0.017   \\
 & & & & & & & & & & & & & & & & & & \\
 HD\,186427   &     -0.170 & -0.836 & -0.111 & 1.098 &  1.369 & 1.455 &  6.963 & 0.015 & 6.231 & 0.010 &   6.3656 & 0.0007  &    -   &   -   &    -   &   -   &  4.651 & 0.016   \\
 & & & & & & & & & & & & & & & & & & \\
 HD\,190360   &     -0.221 & -1.002 & -0.162 & 1.189 &  1.505 & 1.602 &  6.661 & 0.014 & 5.811 & 0.009 &   5.8775 & 0.0006  &    -   &   -   &    -   &   -   &    -   &   -     \\   
 & & & & & & & & & & & & & & & & & & \\
 HD\,207978   &     -0.140 & -0.550 & -0.096 & 0.868 &  1.076 & 1.141 &  6.021 & 0.014 & 5.572 & 0.009 &   5.6173 & 0.0008  &    -   &   -   &    -   &   -   &    -   &   -     \\   

 & & & & & & & & & & & & & & & & & & \\

   \hline  
\end{tabular}
\tablefoot{Adopted bolometric corrections (BC). Hipparcos $H_p$ and Tycho2
       $B_TV_T$ magnitudes for all stars. 2MASS $JHK_S$ only if the quality is
       flagged `A'. The zero-point of these bolometric corrections is set by $M_{bol,\odot}=4.75$. $F_\mathrm{bol}$ for each star is obtained from Eq. (3) of \cite{Casagrande_VandenBerg18a}, adopting L$_{\odot}=3.842 \times 10^{33}$ erg s$^{-1}$.}
\label{bolcor}
  \end{sidewaystable*}
}
%%%

\section{Methods}
Our methodology for delivering the stellar parameters is largely consistent with that in  \citet{karovicova20} so the  results are homogeneous. The connection between the interferometric,
photometric and spectroscopic analysis is  described in detail therein. Here we repeat some key points, and point out differences and updates. 

\subsection{Modelling of limb-darkened angular diameters}
\label{models}

Observations in the first lobe of the visibility function are degenerate between the angular diameter and limb-darkening. A limb-darkened disc appears much the same as a slightly smaller uniformly illuminated disc at these spatial frequencies. To determine the true angular diameters of our targets we therefore rely on stellar atmosphere models to infer the amount of limb-darkening present. 

Limb-darkening laws are used to parametrise the intensity profile with a small number of coefficients. It is common in interferometric studies to use a linear limb-darkening law \citep{Schwarzschild1906}, 
\begin{equation}
\frac{I\left(\mu\right)}{I\left(1\right)}=1-u\left(1-\mu\right), \label{eqn:lin}
\end{equation}
where $\mu = \cos(\gamma)$ and $\gamma$ is the angle between the line of sight and the normal to a given point on the stellar surface, and $u$ is the linear limb-darkening coefficient. However, the linear law does not adequately reproduce the intensity profiles of observations \citep[e.g.][]{klinglesmith70} and of models \citep[e.g.][]{claret11}. For this study, as for our previous studies \citep{karovicova18,karovicova20}, we therefore adopted the four-term non-linear limb-darkening law of \citet{claret00},
\begin{equation}
    \frac{I\left(\mu\right)}{I\left(1\right)} = 1 - \sum_{k}^{4}a_k\left(1-\mu^{k/2}\right),\label{eqn:LD-4Term}
\end{equation}
where $a_k$ is the limb-darkening coefficient.

We derived the limb-darkening coefficients for our targets from the STAGGER grid of ab initio 3D hydrodynamic stellar model atmosphere simulations \citep{magic13}. We use these models because they have been shown to better reproduce the solar limb-darkening than both theoretical and semi-empirical 1D hydrostatic models \citep{pereira13}; therefore, we expect them to give better overall results in general. For each model in the grid and for each of the 38 wavelength channels of PAVO we fitted Equation~\ref{eqn:LD-4Term} to the $\mu$-dependent synthetic fluxes calculated by \citet{magic15}. We then linearly interpolated within this grid to find the limb-darkening coefficients for each star, which are given in Tables A.1.--A.9.,
%~\ref{table:A1}--\ref{table:A8} 
available at the CDS. In the appendix we include one table for one of the stars as an example.

The fringe visibility is related to the source intensity distribution by a Fourier transform. Following the fringe visibility for a generalised polynomial limb-darkening law \citep{quirrenbach96}, the fringe visibility for the four-term non-linear law is
\begin{multline}
    V = \left(\frac{1-\sum a_k}{2}+\sum_k\frac{2a_k}{k+4}\right)^{-1} \times \\
 \left[\left(1-\sum_k a_k\right)\frac{J_1(x)}{x}+\sum_k a_k 2^{k/4}\Gamma\left(\frac{k}{4}+1\right)\frac{J_{k/4+1}(x)}{x^{k/4+1}} \right],
 \end{multline}
where $x=\pi B \theta \lambda^{-1}$, with $B$ the projected baseline, $\theta$ the angular diameter, $\Gamma(z)$ is the gamma function, and $J_n(x)$ is the $n$th-order Bessel function of the first kind. The quantity $B\lambda^{-1}$ is the spatial frequency. We fitted this equation to our observed visibilities to determine the angular diameters of our targets. We determined the uncertainties in our measurements by Monte Carlo simulations that incorporated the uncertainties in the visibility measurements, the wavelength calibration (5\,nm), calibrator sizes (5\%), and the limb-darkening coefficients.
The correlations between wavelength channels are taken into account by bootstrapping.
The raw error bars are scaled by $\chi^{2}$ before the final fitting.
The squared visibilities versus spatial frequencies along with the residuals from the fit are shown in in Figs. \ref{figures1}--\ref{figures5}.
%We adopted the 3D hydrodynamical models that have been shown to better reproduce the solar limb-darkening than both theoretical and semi-empirical 1D hydrostatic models \citep{pereira13}, therefore, they are expected to give better overall results. 
%We updated our analysis in comparison to previous data reduction \citep{karovicova18}, now we are using different limb-darkening coefficients in each of the 38 wavelength channels. Before the data were treated with a single coefficient calculated in R\,band. We have also changed the wavelength range we use in comparison to \citet{white13} and \citet{huber12}. Now the range is 0.63-0.88 microns, though previously the range was 0.65-0.8 microns. 

While our recommended stellar parameters are based on angular diameters determined in this way, for ease of comparison with other studies where the treatment of limb-darkening may be different, we   also calculated angular diameters for a uniformly illuminated disc model, that is, without limb-darkening. Additionally, we   determined the angular diameters for a linearly limb-darkened disc model using coefficients determined from the grid of \citet{claret11}, as had been done for the previous studies using PAVO data that we reanalyse here. These values are given in Table~\ref{table:linear}. We note that the angular diameters are on average 0.8\% smaller than, but still within $1\,\sigma$ of, the values obtained in the previous studies. These differences are due to the updated analysis, as described in Sect.~\ref{sec:observations}. 
The limb-darkened diameters based on the STAGGER-grid are listed in Table~\ref{der_parameters}.

\subsection{Bolometric flux}

The bolometric fluxes and associated uncertainties were derived with the exact same procedure described in \cite{karovicova20}. Briefly, we adopted an iterative procedure to interpolate over the tables of bolometric corrections\footnote{\url{https://github.com/casaluca/bolometric-corrections}} of \cite{Casagrande_VandenBerg14,Casagrande_VandenBerg18a}. We used Hipparcos $H_p$ and Tycho2 $B_TV_T$ magnitudes for all stars, and 2MASS $JHK_S$ only if they had quality flag `A'. Reddening was assumed to be zero for our science targets as they are all located between $\sim$ 10 and 30 pc from us, and thus well within the Local Bubble. 
The adopted bolometric corrections are listed in Table~\ref{bolcor}. 

% FIXME: give the Fbol_sol constant adopted in our calculations

We note that our uncertainties in bolometric fluxes do not take into account possible inaccuracies in model fluxes. Comparison with absolute spectrophotometry indicates that by using multiple bands, bolometric fluxes from the CALSPEC library can be recovered at the one percent level for FG stars \citep{Casagrande_VandenBerg18a}. 
However, our sample comprises cooler stars, for which the performance of our bolometric corrections is much less tested \citep[see e.g. discussions in][]{white18,rains21,tayar20}.
Encouragingly, comparison with the absolute spectrophotometry of a few GK subgiants in \citet{white18} indicates that reliable fluxes can still be obtained from our bolometric corrections. As pointed out earlier, it should also be kept in mind that a given percentage change in bolometric flux carries a four times smaller percentage change in effective temperatures.

     \begin{table}%\small
   \caption{Derived stellar parameters ($T_\mathrm{eff}$, $\log\,g$, $[\mathrm{Fe/H}]$)}    
 %  \centering   
  \begin{tabular}{l l l  r }      
\hline\hline   
 Star &   $T_\mathrm{eff}$  &    $\log\,g$ &     $[\mathrm{Fe/H}]$  \\
    &      (K) & (dex)& (dex) \\ 
  \hline
HD\,131156 & 5545 $\pm$ 92 & 4.561 $\pm$ 0.017 & $-$0.10 $\pm$ 0.04 \\
HD\,146233 & 5819 $\pm$ 31 & 4.437 $\pm$ 0.013 & 0.06 $\pm$ 0.03   \\ 
HD\,152391 & 5380 $\pm$ 45 & 4.486 $\pm$ 0.017 & $-$0.10 $\pm$ 0.06 \\
HD\,173701 & 5295 $\pm$ 46 & 4.426 $\pm$ 0.017 & 0.20 $\pm$ 0.08    \\
HD\,185395 & 6853 $\pm$ 29 & 4.254 $\pm$ 0.010 & 0.06 $\pm$ 0.06    \\
HD\,186408 & 5864 $\pm$ 48 & 4.302 $\pm$ 0.014 & 0.15 $\pm$ 0.05  \\  
HD\,186427 & 5814 $\pm$ 59 & 4.373 $\pm$ 0.015 & 0.12 $\pm$ 0.03    \\
HD\,190360 & 5557 $\pm$ 22 & 4.292 $\pm$ 0.012 & 0.17 $\pm$ 0.04   \\
HD\,207978 & 6403 $\pm$ 30 & 4.070 $\pm$ 0.035 & $-$0.53 $\pm$ 0.07   \\
  \hline                                 
\end{tabular}
\label{der_parameters2}
  \end{table}

\subsection{Spectroscopic analysis}
We measured a set of unblended \ion{Fe}i and \ion{Fe}{ii} lines from high-resolution $R \approx 42,000$ spectra from the ELODIE archive \citep{Moultaka2004}. 
We used a pipeline based on the Spectroscopy Made Easy (SME) code \citep{piskunov_spectroscopy_2017}, together with MARCS model atmospheres \citep{gustafsson_grid_2008} and non-LTE corrections from \citet{amarsi_non-lte_2016-1}.

We performed a differential abundance analysis by a comparison to measurements on a solar spectrum recorded with the same spectrograph from reflected light off the Moon. These differential measurements remove zero-level uncertainties in the reference oscillator strengths, and ensure that metallicity measurements are relative to the Sun with no significant zero-point uncertainty. 
Our iron abundance estimates are based on an outlier-resistant mean with $3 \sigma$ clipping. We also estimate abundances separately for \ion{Fe}i and \ion{Fe}{ii}; comparing these two measurements offers an independent check on the accuracy of our stellar parameters as systematic errors are expected to yield deviations from ionisation equilibrium.
In addition to our statistical uncertainties, we also estimated the  systematic errors on the metallicity. These estimates were computed from the effect of uncertainties in each of the stellar parameters on the metallicity measurements, and then combined in quadrature.

%%%%

\subsection{Stellar evolution models}
We determined stellar masses using the Dartmouth stellar evolution tracks \citep{Dotter2008}, which cover a wide range of ages and metallicities. We used solar-scaled models with $\rm [\alpha/Fe] = 0$ since the stars investigated in this work are thin-disc stars with no discernible $\alpha$-enhancement at low metallicity.
We performed the fitting using the ELLI package\footnote{ \url{https://github.com/dotbot2000/elli}} \citep{Lin18}, which uses a Bayesian framework to estimate mass and age from our estimates of  $T_\mathrm{eff}$, $\log L/L_\odot$, and $[{\rm Fe }/{\rm H}]$. Errors on these parameters are assumed to be symmetric and independent. This code samples the posterior distribution using a Markov chain Monte Carlo (MCMC) method; we estimate the mass and its uncertainty from the mean and standard deviation of this distribution.
The surface gravity is computed directly from this estimate on the form
\begin{equation}
\log g = \log \frac {G M}{R^2} = \log \frac {4 G M \varpi^2}{\theta^2},
\label{eq:logg}
\end{equation}
where $G$ is the gravitational constant and $\varpi$ the parallax.

%%%%%

   \begin{table}%\small
%   \caption{Contributions to the uncertainties in $T_\mathrm{eff}$} 
                \caption{{Uncertainties in $T_\mathrm{eff}$ and how they propagate
       from the underlying measurements.}}
  \centering   
  \begin{tabular}{l l r r r r }      
\hline\hline   
 
% Star &   $T_\mathrm{eff}$  & e$T_\mathrm{eff}$   & e$F_\mathrm{bol}$\,=\,0  & e$\theta_{LD}$\,=\,0  \\
%    &      (K) & (K)& (K) & (K)\\ 
         Star &   $T_\mathrm{eff}$  & e$T_\mathrm{eff}$  & e$T_\mathrm{eff}$   & e$F_\mathrm{bol}$ $^{a}$ & e$\varTheta_{LD}$ $^{b}$ \\
    &      (K) & (K)& (\%) &(K) & (K)\\ 
  \hline 
HD\,131156 & 5545  & 92  & 1.7 & \bf{89} & 22 \\
HD\,146233 & 5819  & 31  & 0.5 & 5       & \bf{31} \\
HD\,152391 & 5380  & 45  & 0.8 & 22      & \bf{39} \\
HD\,173701 & 5295  & 46  & 0.9 & 22      & \bf{40} \\
HD\,185395 & 6853  & 29  & 0.4 & 8       & \bf{28} \\
HD\,186408 & 5864  & 48  & 0.8 & \bf{42} & 22 \\
HD\,186427 & 5814  & 59  & 1.0 & \bf{54} & 24 \\
HD\,190360 & 5557  & 22  & 0.4 & 8       & \bf{21} \\
HD\,207978 & 6403  & 30  & 0.5 & 6       & \bf{30} \\
   \hline  
\end{tabular}
\tablefoot{\tablefoottext{a}{The uncertainty contributions from the bolometric flux if the
$\varTheta_{LD}$ uncertainties are set to 0.
$^{(b)}$  The uncertainties arising entirely from the angular diameter measurements if the $F_\mathrm{bol}$
 uncertainties are set to 0. The dominating uncertainty is highlighted in boldface.}}
\label{errors}
  \end{table}

\section{Results and discussion}
\label{results}
\subsection{Recommended stellar parameters}

We present fundamental stellar parameters and angular diameters 
for a set of nine dwarf stars: HD\,131156 ($\xi$\,Boo), HD\,146233 (18\,Sco), HD\,152391, 
HD\,173701, HD\,185395 ($\theta$\,Cyg), HD\,186408 (16\,Cyg\,A), HD\,186427 (16\,Cyg\,B), HD\,190360, and HD\,207978 (15\,Peg). 
We present the stars
as a new set of benchmark stars.
We estimate radius and mass from measurements of $\theta_{LD}$, $F_\mathrm{bol}$, and parallax
for all nine stars.
All these values, along with luminosity, 
are summarised in Table~\ref{der_parameters}.
The final fundamental stellar parameters of  $T_\mathrm{eff}$, $\log\,g$, and $[\mathrm{Fe/H}]$ are presented in Table~\ref{der_parameters2}, and are discussed below.

\subsection{Uncertainties}
\label{uncertainties}

The final uncertainties in $T_\mathrm{eff}$ 
are due, independently, to the uncertainties in the bolometric flux and to the uncertainties in the angular diameter.
The contributions from each are shown in Table~\ref{errors}, where they have been computed by artificially setting the uncertainties from the other measurement to zero. For clarity the dominating uncertainty is highlighted in boldface.

It should be noted that in the final $T_{\mathrm{eff}}$ error estimates,
we propagate the statistical measurement uncertainties in 
$\log\,g$ and $\mathrm{[Fe/H]}$ from the
isochrone fitting and spectroscopic analysis, which were folded into the uncertainties
in the angular diameters. The median uncertainties in 
$\log\,g$ and $\mathrm{[Fe/H]}$ across our sample of
stars are 0.015\,dex and 0.05\,dex, respectively (Table~\ref{der_parameters2}).

The uncertainties in $T_\text{eff}$ are less than 50-60~K (or less than\,1\,\%) for all but two stars in our sample. For one of these the uncertainty is equal to 1\%.
For the remaining star the uncertainty in $T_\text{eff}$
is higher than 1\%, however, still low (1.7\%).
The uncertainty for these two stars is dominated by errors in the bolometric flux. 
Errors in $T_\text{eff}$ resulting from uncertainties in the limb-darkened angular diameters are at most 40\,K, with a median of just 29\,K (0.5\,\%). 
The only star in our dwarf sample
for which we report $T_\mathrm{eff}$ uncertainties
larger than 1\% (1.7\,\%) is HD\,131156.
As already mentioned, the $T_\mathrm{eff}$ uncertainty is dominated by the uncertainty in the bolometric flux; therefore, 
the $F_\mathrm{bol}$ value for this star requires refinement before the star fully meets the set requirements for the $T_\mathrm{eff}$ precision. 
As mentioned previously, the desired precision
requested by the spectroscopic
teams such as  {\it Gaia}-ESO or GALAH is around 1\% (or around 40-60~K).
%%%

\subsection{Comparison with angular diameter values in the literature}
\label{comparison}

\begin{figure}
        \includegraphics[width=\columnwidth]{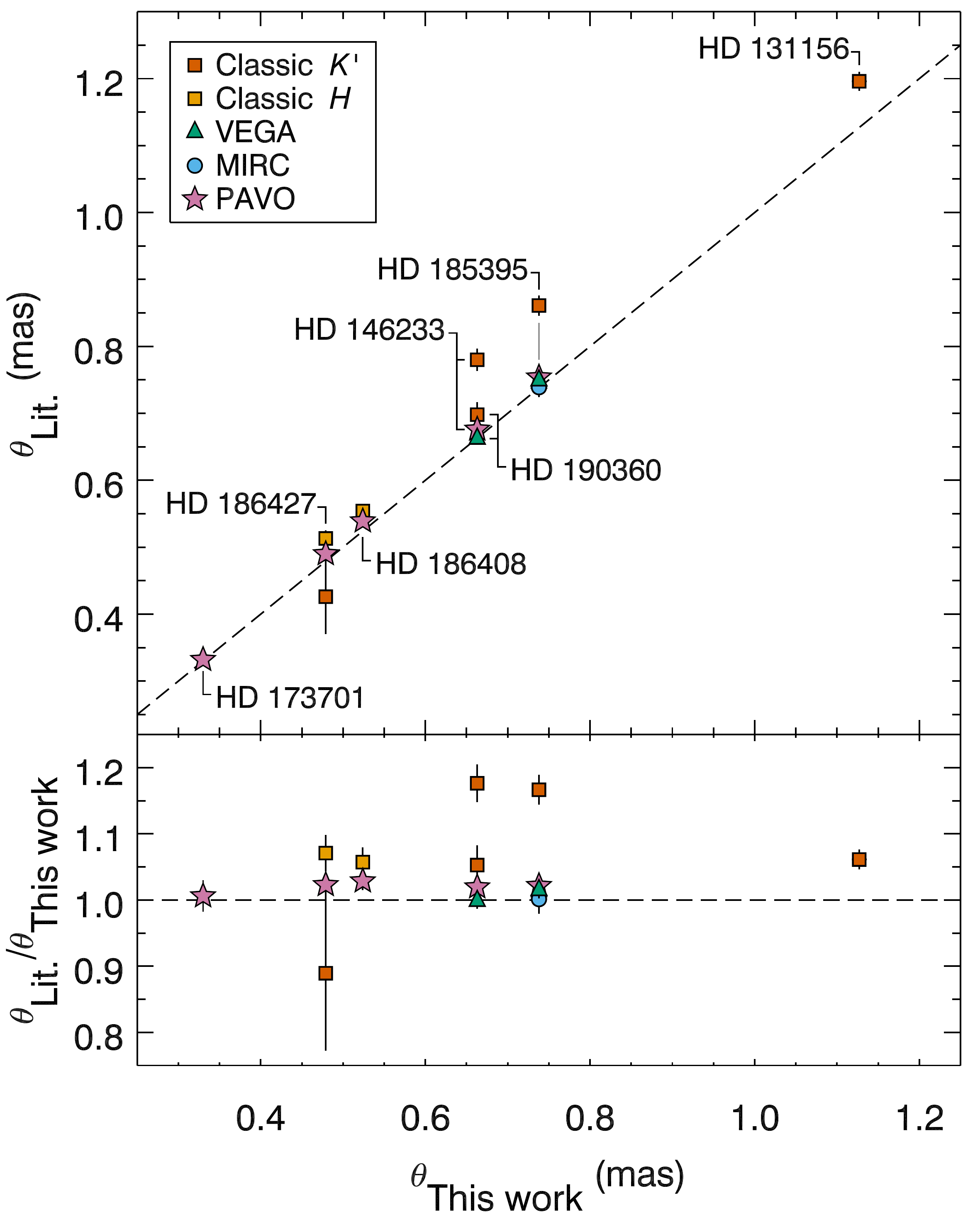}
    \caption{Comparison of limb-darkened angular diameters from the literature with measurements from this work. Symbols correspond to the beam combiner used for the literature measurement: CHARA Classic in the $K'$ and $H$ band (orange and yellow squares, respectively), VEGA in the $H$ band (green triangles), MIRC in the $H$ band (light blue circle), and PAVO in visible (pink stars).}
    \label{fig:compare}
\end{figure}

In total, seven of the stars have been previously interferometrically 
observed. We present the first published angular diameters for 
two stars: HD\,152391 and HD\,207978.
For some of the stars we conducted new observations and for all the stars we carried out fresh data reduction and analysis of the data. For all the stars we applied our updated analysis and updated treatment of limb-darkening modelling.
Table~\ref{prediam} lists our measurements of angular diameters $\theta_{LD}$ together with values reported in the literature. We compare these values in Fig.~\ref{fig:compare}.

   \begin{table*}%\small
   \caption{Previous Angular Diameters}    
  \centering   
  \begin{tabular}{l c c l l } 
  \hline
  \hline
Star              &             Our value                &      Literature value    &      Reference                &      CHARA\\
&{(mas)}&{(mas)}&&Instrument\\
\hline
HD\,131156                &     1.124±0.009    &&&\\
&&      1.196±0.014                     &      \citet{Boyajian12a}      &       Classic $K'$\\
\hline
%  &  &  &   &   &        \\
HD\,146233        &             0.663±0.007    &&&\\
& &     0.780±0.017                     &      \citet{Boyajian12a}      &       Classic $K'$\\
                          &              &              0.676±0.006                     &       \citet{Bazot11}  &      PAVO\\
%   &  &  &   &   &  \\
\hline
HD\,152391                &     0.477±0.007     &  &  &                \\                                      
%    &  &  &   &   & \\
\hline
HD\,173701                &     0.329±0.005    &&&\\
&&      0.332±0.006                     &      \citet{huber12}  &      PAVO\\
%   &  &  &   &   &   \\
\hline
HD\,185395                &     0.737±0.006    &&&\\
&&      0.861±0.015             &              \citet{Boyajian12a}      &       Classic $K'$\\
% &                               &                     0.760±0.003             &               Ligi+ 2012       &      VEGA\\
&                                 &                     0.754±0.009             &               \citet{white13}  &      PAVO\\
&                                 &                     0.739±0.015             &               \citet{white13}  &      MIRC\\
&                         &                             0.749±0.007             &               \citet{Ligi16}   &      VEGA\\
%  &  &  &   &   & \\
\hline
HD\,186408                &     0.525±0.004    &&&\\
&&      0.554±0.011             &              \citet{Boyajian13}       &       Classic $H$\\
                          &                      &      0.539±0.006             &               \citet{white13}          & PAVO\\
%  &  &  &   &   & \\
\hline
HD\,186427                &     0.479±0.004    &&&\\
&&      0.426±0.056             &              \citet{Baines08}         &       Classic $K'$\\
                          &                      &      0.513±0.012             &               \citet{Boyajian13}       &      Classic $H$\\
                          &                      &      0.490±0.006             &               \citet{white13}          & PAVO\\
%  &  &  &   &   & \\
\hline
HD\,190360                &     0.663±0.005    &&&\\
& &     0.698±0.019     &                      \citet{Baines08}         &       Classic $K'$\\
                          &                      & 0.662±0.006                 &         \citet{Ligi16}   &      VEGA\\
%  &  &  &   &   & \\
\hline
HD\,207978                &     0.542±0.005             &  &  &        \\
   \hline  
   \label{prediam}
\end{tabular}
  \end{table*}

{\bf HD\,131156 ($\xi$\,Boo)} We observed this star and measured the angular diameter as
$\theta_{LD}=1.124\pm0.009\,\text{mas}$.
This star was previously interferometrically observed by 
\citet{Boyajian12a} using the Classic instrument in the $K'$ band
at the CHARA array who reported
$\theta_{LD}=1.196\pm0.014\,\text{mas}$.
We note that the difference of the $\theta_{LD}$ value in comparison 
with the previous study 
is almost 4 $\sigma$ over the quoted uncertainties.
We discuss possible reasons for the difference between our measurement using the PAVO instrument and the measurement from the Classic instrument below. This star was suggested as a possible benchmark star in \citet{heiter15}; therefore, we included it in our observing sample.

{\bf HD\,146233 (18\,Sco)} We measured angular diameter of this star as
$\theta_{LD}=0.663\pm0.005\,\text{mas}$.
18\,Sco was previously observed by \citet{Bazot11} with the same interferometric 
instrument, PAVO. However, the data were collected 
on a single night and so are potentially susceptible to calibration errors. 
The published angular diameter from that night is
$\theta_{LD}=0.676\pm0.006\,\text{mas}$.
18\,Sco is a Gaia-ESO benchmark and the value of $\theta_{LD}$ reported by \citet{Bazot11} was adopted by the Gaia-ESO survey \citep{jofre15}.
The values of $\theta_{LD}$ differ by 1.4$\sigma$.
This difference is due, in part, to the additional observations included in our measurement, but also because of the different limb-darkening correction.
% we have used based on the 3D STAGGER grid is weaker than the correction used by \citet{Bazot11} drived from 1D ATLAS models by \citet{claret00}.
Additional interferometric observations of this star were
conducted by \citet{Boyajian12a}, who observed the star
using the Classic instrument at the CHARA array and
reported $\theta_{LD}=0.780\pm0.017\,\text{mas}$.
Here the difference between the values is 
over 6 $\sigma$. We again discuss the possible reasons in more detail below.

{\bf HD\,152391} This star was interferometrically
observed for the first time.
We report the angular diameter of
$\theta_{LD}=0.477\pm0.007\,\text{mas}$.

{\bf HD\,173701} For this star we report
$\theta_{LD}=0.329\pm0.005\,\text{mas}$.
The data were previously analysed by \citet{huber12},
who reported $\theta_{LD}=0.332\pm0.006\,\text{mas}$. \citet{huber12} fitted a linearly limb-darkened disc model to the data, determining the limb-darkening coefficient from the grid calculated by \citet{claret11} from 1D ATLAS models.
For consistency, we carried out new data reduction and re-analysed the data with limb-darkening coefficients from 3D model atmospheres and using a higher-order limb-darkening model.
In this case, the angular diameters are in agreement.
This star was suggested as a possible benchmark star in \citet{heiter15}.

{\bf HD\,185395 ($\theta$\,Cyg)} This star was previously observed several 
times with various interferometric instruments. 
For this star we report
$\theta_{LD}=0.737\pm0.006\,\text{mas}$.
We re-visited data observed by \citet{white13} who found a value of $\theta_{LD}=0.754\pm0.009\,\text{mas}$.
The differences between our value of $\theta_{LD}$ and the value determined by \citet{white13} can be explained by a combination of our new reduction of the data and the different treatment of limb-darkening.
We note that our value is in agreement with the measurement made by \citet{white13} using the MIRC beam combiner at CHARA ($\theta_{LD}=0.739\pm0.015\,\text{mas}$), which operates in the $H$ band where limb-darkening is weaker. Our value is slightly smaller than the value obtained at visible wavelengths with the VEGA beam combiner at CHARA by \citet{Ligi16}, who found $\theta_{LD}=0.749\pm0.007\,\text{mas}$. All these values are significantly smaller than the value obtained with the Classic in the $K'$ band of 
$\theta_{LD}=0.861\pm0.015\,\text{mas}$ \citep{Boyajian12a}. This star was also suggested as a possible benchmark star.

{\bf HD\,186408 (16\,Cyg\,A)} For this star we report 
$\theta_{LD}=0.525\pm0.004\,\text{mas}$.
Here, we carried out a new data reduction and re-analysis of data observed with the PAVO instrument by \citet{white13},
who reported $\theta_{LD}=0.539\pm0.006\,\text{mas}$.
Again, this difference can be attributed to our new reduction and the updated limb darkening treatment.
We discuss the matter in more detail below.
\citet{Boyajian13} published observations of this star
using the Classic instrument in the $H$ band and found a larger value,
$\theta_{LD}=0.554\pm0.011\,\text{mas}$. This star is listed as a candidate for a benchmark \citep{heiter15}.

{\bf HD\,186427 (16\,Cyg\,B)} For this star we report
$\theta_{LD}=0.479\pm0.004\,\text{mas}$.
We re-visited PAVO data observed by \citet{white13}.
The previous reported value of $\theta_{LD}$ was $\theta_{LD}=0.490\pm0.006\,\text{mas}$.
CHARA Classic observations have been made in the $K'$ band by \citet{Baines08} and in the $H$ band by \citet{Boyajian13}; the authors
reported angular diameters of 
$\theta_{LD}=0.426\pm0.056\,\text{mas}$
and $\theta_{LD}=0.513\pm0.012\,\text{mas}$, respectively.
Since the star was suggested as a possible candidate as a benchmark star, for consistency the data
was freshly reduced, re-analysed, and the limb-darkening coefficients from 3D
model atmospheres using a higher-order limb-darkening model
were applied.

{\bf HD\,190360} We observed the star and measured angular diameter as
$\theta_{LD}=0.663\pm0.005\,\text{mas}$. Our value agrees with the value, $\theta_{LD}=0.662\pm0.006\,\text{mas}$, found with the VEGA instrument by \citet{Ligi16}.
This star was also previously observed 
by \citet{Baines08} using the CHARA Classic instrument in the $K'$ band, who found $\theta_{LD}=0.698\pm0.019\,\text{mas}$. Possible reasons for this difference are discussed below.

{\bf HD\,207978 (15\,Peg)} The last star in our dwarf sample has not been previously
observed interferometrically.
We determined the angular diameter to be
$\theta_{LD}=0.542\pm0.005\,\text{mas}$.\\

As we have noted, the differences between previous PAVO diameters and our values based on the same data are attributable to a combination of our fresh data reduction and changes to the limb-darkening correction. Both of these changes contribute almost equally to our diameters being an average of 2\% smaller than the previous PAVO values.

Most of the changes in our new data reduction had only a small impact on the result. These changes include making different choices around outlier rejection based on S/N and other metrics, new calculations of calibrator sizes, and using more of the observed wavelength channels. The most significant impact results from our choice not to use the $t_0$ correction, as described in Sect.~\ref{sec:observations}. The effect of these data reduction changes is reflected in the limb-darkened angular diameters given in Table~\ref{table:linear}, which used linear limb-darkening coefficients determined in the same way as in the previous studies.

The change to the limb-darkening treatment makes up the remainder of the difference. The previous PAVO studies used linear limb-darkening coefficients calculated from 1D ATLAS model atmospheres in Cousins $R$ band, either by \citet{claret00} or \citet{claret11}. Two different methods of determining the linear coefficient were used by \citet{claret11}, a simple least-squares (LS) fit to the intensity profile, and a flux-conserving (FC) fit. The previous PAVO studies adopted the average of the values determined by these methods, and used the difference as an estimate of the uncertainty. The $R$ band value was used across all wavelength channels between 650 and 800\,nm.

\begin{figure}
        \includegraphics[width=\columnwidth]{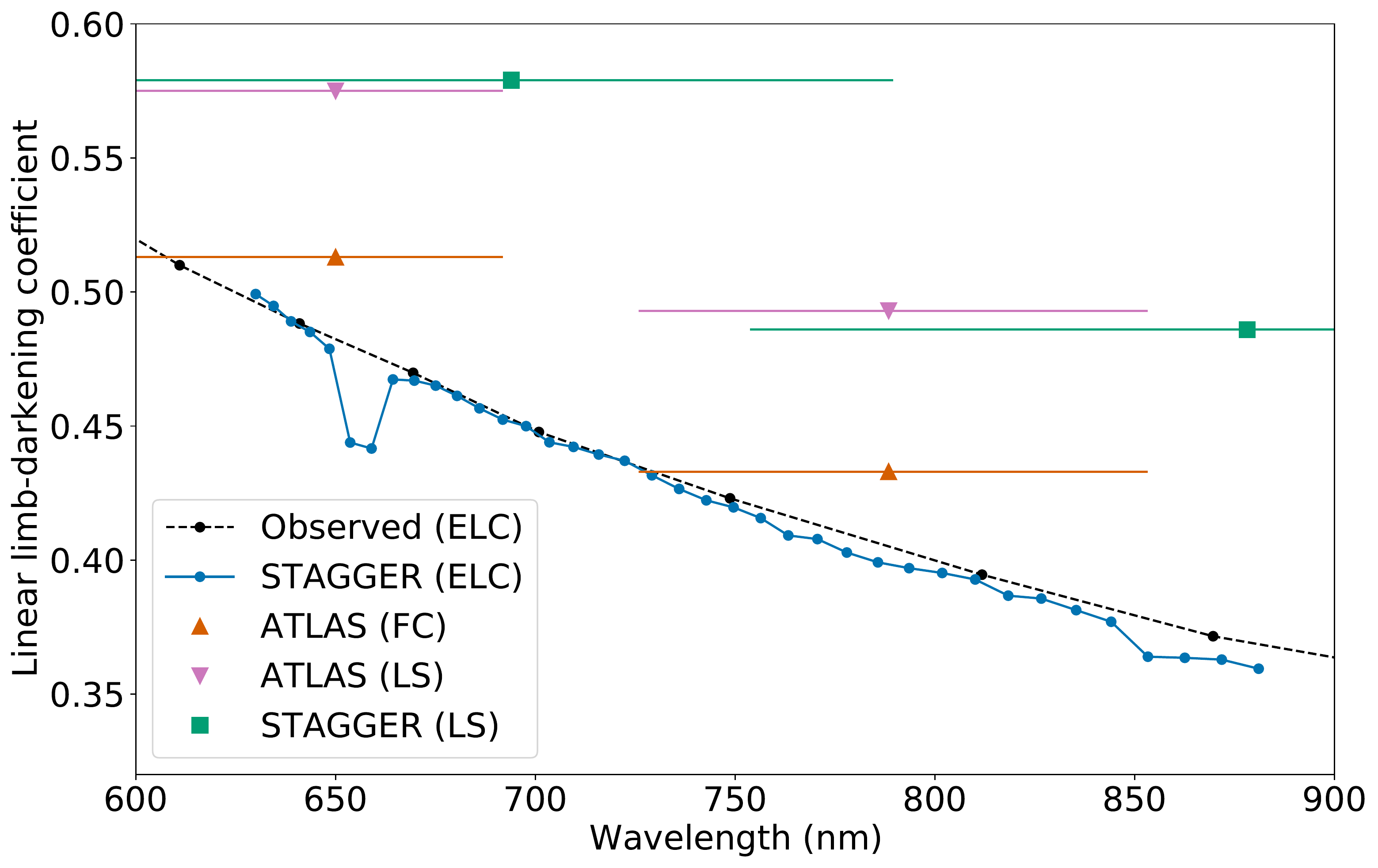}
    \caption{Solar linear limb-darkening coefficients as a function of wavelength. Black points, joined by a dashed line to guide the eye, show the equivalent linear coefficients (ELCs) calculated from the observations of \citet{NeckelLabs94}. Blue points, joined by a solid blue line, show the ELCs calculated from the 3D STAGGER solar model \citep{magic15} in each of the 38 wavelength channels of PAVO. Triangles show linear coefficients calculated by \citet{claret11} from 1D ATLAS models in Cousins $R$ and $I$ bands. They used two different methods, flux-conserving (FC) and least-squares (LS) fits  to the models, indicated by upward orange triangles and downward pink triangles, respectively. Green squares show the linear coefficients calculated using a LS fit from the STAGGER model in Johnson $R$ and $I$ bands by \citet{magic15}. The error bars show the full width at half maximum of the respective bandpasses. See text for a discussion of the different methods of determining the coefficients.
  }
    \label{fig:compare-ld}
\end{figure}

As noted in Sect.~\ref{models}, the linear limb-darkening law is a poor fit to both observations and models of centre-to-limb intensity profiles. This is so because the intensity drops sharply close to the limb. As a result, a simple linear LS fit to the intensity profile will produce a stronger limb-darkening across the vast majority of the stellar disc than is appropriate. The FC fit of \citet{claret11} takes account of this by altering the slope of the fit to ensure that the integrated flux over the stellar disc is conserved. 

An alternate way to determine an appropriate linear coefficient was recently used by \citet{rains20}. Their equivalent linear coefficient (ELC)  produces the same first sidelobe height in the interferometric visibility function as higher-order models or observations. 

The differences between these coefficients are illustrated by Fig.~\ref{fig:compare-ld}, where we show model and observed linear limb-darkening coefficients for the Sun as a function of wavelength. The ELCs that correspond to the four-term non-linear  limb-darkening law coefficients we  use in this work agree well with ELCs that match the solar observations of \citet{NeckelLabs94}. By comparison, the linear coefficients determined by \citet{claret11} from 1D ATLAS models and by \citet{magic15} from the 3D STAGGER grid imply stronger limb-darkening, with the coefficients determined by a LS fit giving the worst results. Additionally, the previous practice of applying the $R$ band value to all wavelength channels also resulted in a bias towards stronger limb-darkening. The previous studies therefore used limb-darkening corrections that were too strong, resulting in angular diameters that were too large.

Even larger differences are found between our measurements and several of those made with the Classic beam combiner, particularly in the $K'$ band. Such differences have been noted previously \citep[e.g.][]{Casagrande14,karovicova18,white18}, and can arise from several factors, including the target star not being resolved well enough. 

\subsection{Spectroscopic analysis}
\begin{figure}
\includegraphics[width=\hsize]{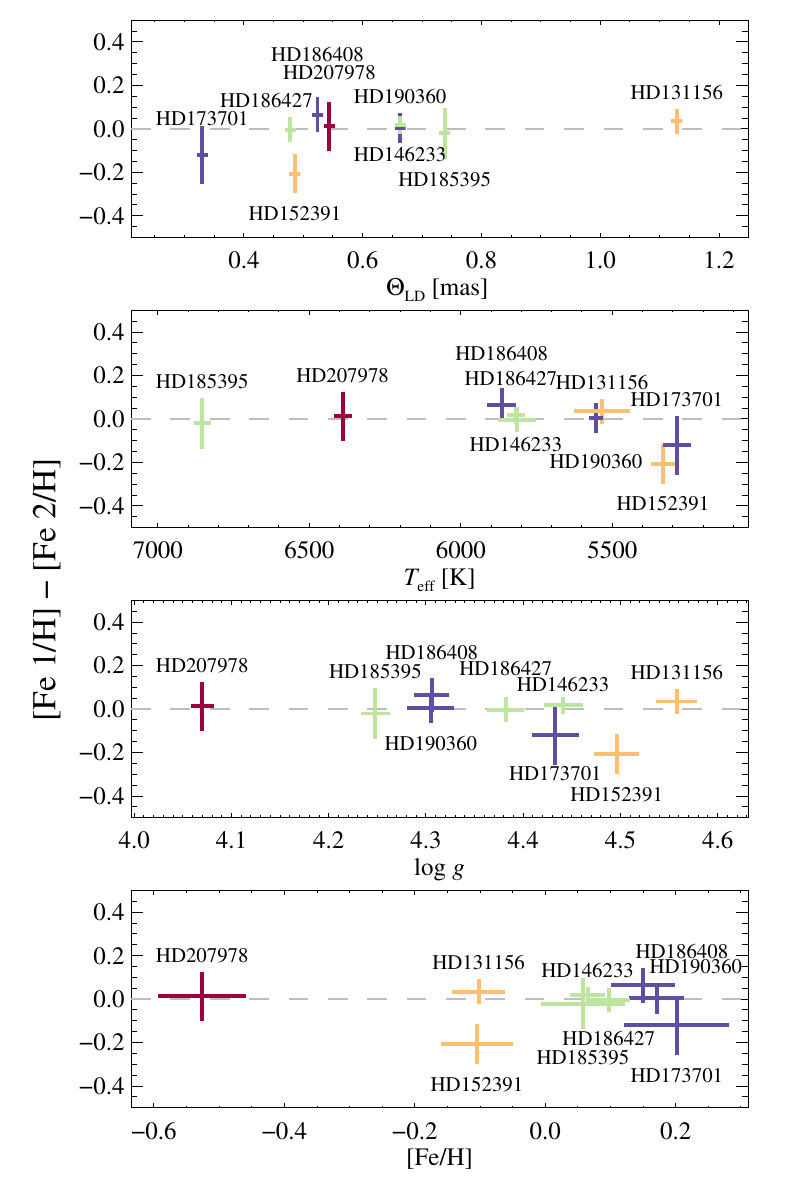}
\caption{Deviations from ionisation equilibrium (i.e. the difference between the abundances determined from lines of neutral and ionised iron) as a function of the measured stellar parameters. Vertical and horizontal lines represent the combined uncertainties from the two measurements. Stars are colour-coded according to metallicity, as in  Fig.~\ref{fig:HRdiag}.}
\label{fig:ionisation_equilibrium}
\end{figure}

We show in Fig.~\ref{fig:ionisation_equilibrium} that our abundance determinations for lines of neutral and singly ionised iron are in good agreement: the median and median absolute deviation are $0.00 \pm 0.02$\,dex.
As errors in stellar parameters have different impact on lines of neutral and singly ionised iron, these measurements offer an independent test of the quality of our stellar parameters. We estimate that an error in $T_\text{eff}$ of 100\,K would affect the abundances measured from neutral iron lines compared to lines ionised by 0.1\,dex; conversely, an error in $\log\,g$ of 0.10\,dex would have a corresponding effect of 0.05\,dex. 

We find that only three out of nine stars exhibit a deviation from ionisation equilibrium greater than their estimated measurement errors, and only significantly so for two stars. We note that these are also the coolest stars in our sample, indicating that the error may be related to our spectroscopic modelling rather than potential errors in the stellar parameters. Crucially, there is no apparent correlation with the angular diameters.

\subsection{Comparison with asteroseismology}
Stellar radii may also be precisely determined from asteroseismic measurements. Such measurements exist for several of our targets; here we briefly compare asteroseismic radii for these stars with our values from interferometry.

The most direct method of determining radii from asteroseismology is to exploit scaling relations for two global asteroseismic parameters, the frequency of maximum power, $\nu_\mathrm{max}$, and the large frequency separation, $\Delta\nu$ \citep[e.g.][]{kallinger10}. The value of $\nu_\mathrm{max}$ is empirically observed to scale as \citep{brown91,kjeldsen95}
\begin{equation}
    \frac{\nu_\mathrm{max}}{\nu_\mathrm{max,\odot}} = \left(\frac{M}{\mathrm{M_\odot}}\right)\left(\frac{R}{\mathrm{R_\odot}}\right)^{-2}\left(\frac{T_\mathrm{eff}}{\mathrm{T_{eff,\odot}}}\right)^{-1/2},
\end{equation}
and $\Delta\nu$ scales as \citep{ulrich86}
\begin{equation}
    \frac{\Delta\nu}{\Delta\nu_{\odot}} = \left(\frac{M}{\mathrm{M_\odot}}\right)^{1/2}\left(\frac{R}{\mathrm{R_\odot}}\right)^{-3/2}.
\end{equation}
By measuring $\Delta\nu$ and $\nu_\mathrm{max}$, and given $T_\mathrm{eff}$, combining these two relations directly gives the mass and radius. A more rigorous method is to model the stars in detail using individual oscillation frequencies and other measurements as observational constraints. Detailed modelling allows  other stellar properties to be constrained, such as age, in addition to the mass and radius.

Four of our targets were observed during the original Kepler mission. Of these, three (HD\,173701, HD\,186408, and HD\,186427) are included in the Kepler LEGACY sample \citep{Lund17, SilvaAguirre17}. We used the measurements of $\nu_\mathrm{max}$ and $\Delta\nu$ by \citet{Lund17} for these stars to calculate scaling relation radii. Detailed asteroseismic modelling for the LEGACY sample was conducted by \citet{SilvaAguirre17} using several modelling pipelines. For simplicity, here we only use the values derived from the BASTA pipeline to illustrate typical modelling results. 

The fourth Kepler star, HD\,185395, was studied by \citet{Guzik16}. We use the $\nu_\mathrm{max}$ and $\Delta\nu$ values they measured to determine the scaling relation radius. \citet{Guzik16} used the interferometric radius as a constraint, along with the oscillation frequencies, in their detailed modelling of this star. While this means that a fair comparison of the modelled radius with the interferometric radius cannot be made, we include their modelled radius determined from YREC models \citep{Demarque08} to illustrate the stark difference between the scaling relation and modelled radius for this star.

\begin{figure}
        \includegraphics[width=\columnwidth]{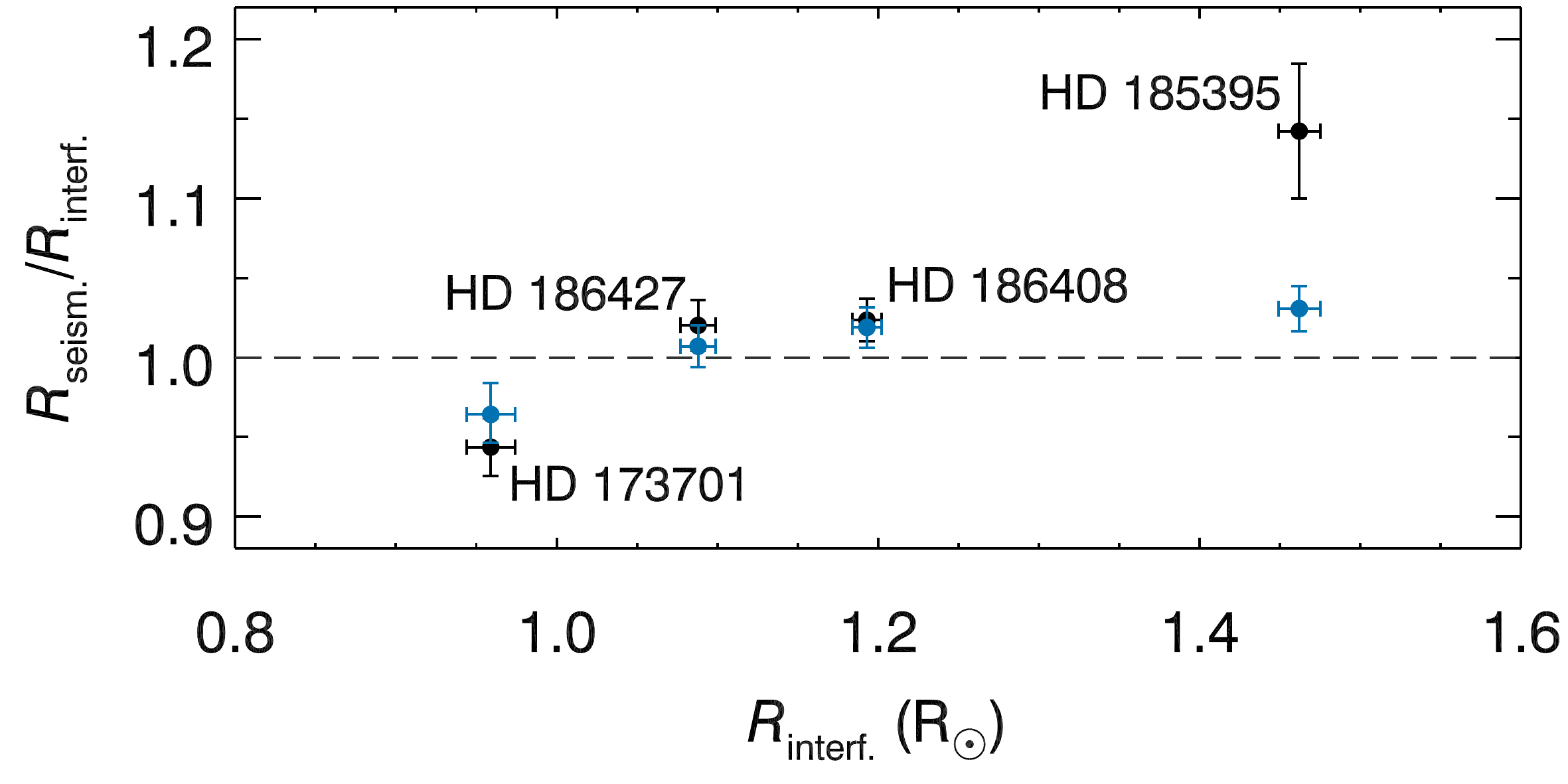}
    \caption{Comparison of radii determined from asteroseismic scaling relations (black) and modelling (blue) with interferometric radii from this work.}
    \label{fig:seismic}
\end{figure}

Figure \ref{fig:seismic} compares the scaling relation and modelling asteroseismic radii with our interferometric radii. As can be expected, the radii determined from the scaling relations do not agree as closely with the interferometric radii as the radii from detailed modelling. The biggest disagreement between interferometry and the scaling relations can be seen for HD\,185395. The apparent failure of the scaling relation in this case can be attributed to it being applied to a star that is more than 1000\,K hotter than the Sun. \citet{White11} highlighted that the $\Delta\nu$ scaling relation, while working well in models,  requires a temperature-dependent correction when applying it to stars significantly different to the Sun. Applying such a correction here would substantially improve the agreement for this star.

In contrast, the agreement between the modelled radii and the interferometric radii is generally good for all stars, although differences remain. \citet{huber12} were unable to reconcile differences between their interferometric and asteroseismic observations, and stellar models for HD\,173701. They noted that a higher effective temperature would help to bring better agreement. Despite our revised angular diameter being slightly smaller than the value determined by \citet{huber12}, we also find a lower bolometric flux than they adopted,  such that the effective temperature remains unchanged. Tension between asteroseismic and interferometric results such as this, however, may be of value for calibrating the  free parameters of stellar models, such as the mixing length parameter \citep[e.g.][]{hjoerringgaard17,Joyce18}.

Finally, we note that HD\,146233 has also been observed asteroseismically by \citet{Bazot11}. However, because a measurement of $\nu_\mathrm{max}$ was not reported by \citet{Bazot11} we have not been able to calculate a scaling relation radius to compare with our measurement. Modelling of this star by \citet{Bazot18} also used the interferometric measurement as a constraint, and subsequently did not report radius as an output of the modelling.

\section{Conclusions}
The goal of this study is to extend a sample of 
stars with highly accurate and reliable fundamental 
stellar parameters that are used as benchmark stars
for large stellar surveys.
This is the second in a series of papers with this scientific goal.
Here we determined fundamental 
stellar parameters of nine dwarf benchmark stars 
HD\,131156 ($\xi$\,Boo\,A), HD\,146233 (18\,Sco), HD\,152391, HD\,173701, HD\,185395 ($\theta$\,Cyg), HD\,186408 (16\,Cyg\,A), HD\,186427 (16\,Cyg\,B), HD\,190360, and HD\,207978 (15\,Peg).

We observed the stars using the high angular resolution instrument PAVO at the CHARA array in order to measure the angular diameters of the stars.
We observed the stars
over various nights and with various baseline
configurations with the aim of resolving the targets close to the first null of the visibility
curve. 
We estimated the limb-darkening diameters using the 3D radiation-hydrodynamical 
model atmospheres in the STAGGER-grid.
We determined the $T_\mathrm{eff}$ directly from from the
Stefan-Boltzmann relation together with photometric modelling of bolometric flux. 
Bolometric fluxes were computed from
multi-band photometry, interpolating iteratively on a grid of synthetic stellar fluxes to ensure consistency with the final adopted
stellar parameters.

Based on high-resolution spectroscopy,
we determined $\rm [Fe/H]$, and we used isochrone fitting to derive 
mass and parallax measurements to constrain the absolute
luminosity. After iterative refinement we derived the final fundamental parameters of $T_\mathrm{eff}$, $\log\,g$, and $\mathrm{[Fe/H]}$. 
One of the stars from the sample HD\,146233 (18\,Sco)
is listed as a Gaia-ESO benchmark star.
There has been a strong disagreement between two interferometric values determined by \citet{Bazot11} and \citet{Boyajian12a}. We resolved the discrepancies and
updated the fundamental stellar parameters
%determined in our study 
for this important solar twin.
We also determined stellar parameters for
five stars previously proposed as benchmarks in \citet{heiter15}. 

For eight stars we reached the required precision of $\lessapprox$\,1\% in the $T_\mathrm{eff}$. 
Only one star (HD\,131156) showed a somewhat larger $T_\mathrm{eff}$ uncertainty, which is still less than 2\,\%. 
The $T_\mathrm{eff}$ uncertainty for this star is dominated by errors in the bolometric flux.
For surface gravity $\log\,g$ we reached a median precision of just 0.015\,dex %??????
%(assuming 10\% uncertainty in mass, 1\% in parallax), 
and 0.05\,dex in metallicity $\mathrm{[Fe/H]}$.
We showed that the updated limb-darkening approach gives slightly different results in comparison to previous interferometric measurements derived using the same instrument, justifying the revisit of previous measurements.

The following paper in this series will extend our sample to giants in the metal-rich range.
These benchmark stars will be used
as a proper scale for stellar models
applied to spectroscopic surveys. Therefore,
the precision we have been able to achieve is crucial for the correct determination of the 
atmospheric parameters of the stars observed by the surveys
where the scaled stellar models are applied.
Properly scaled models will allow us to
better understand the physics of stars
in our Galaxy.\\

\begin{acknowledgements}
      I.K. acknowledges the German
      \emph{Deut\-sche For\-schungs\-ge\-mein\-schaft, DFG\/} project
      number KA4055 and the European Science Foundation - GREAT Gaia Research for European Astronomy Training.
      This work is based upon observations obtained with the Georgia State University Center for High Angular Resolution Astronomy Array at Mount Wilson Observatory. The CHARA Array is supported by the National Science Foundation under Grants No. AST-1211929 and AST-1411654. Institutional support has been provided from the GSU College of Arts and Sciences and the GSU Office of the Vice President for Research and Economic Development.
      Funding for the Stellar Astrophysics Centre is provided by The Danish National Research Foundation.
      L.C. is the recipient of the ARC Future Fellowship FT160100402. T.N. acknowledges funding from the Australian Research Council (grant DP150100250).
      Parts of this research were conducted by the Australian Research Council Centre of Excellence for All Sky Astrophysics in 3 Dimensions (ASTRO 3D), through project number CE170100013.
      D.H. acknowledges support from the Alfred P. Sloan Foundation, the National Aeronautics and Space Administration (80NSSC19K0379), and the National Science Foundation (AST-1717000).
      This work is based on spectral data retrieved from the ELODIE archive at Observatoire de Haute-Provence (OHP). %, and on observations made with the Nordic Optical Telescope, operated by the Nordic Optical Telescope Scientific Association at the Observatorio del Roque de los Muchachos, La Palma, Spain, of the Instituto de Astrofisica de Canarias. % required for the ELODIE and NOT/FIES spectra
\end{acknowledgements}

\bibliographystyle{aa} %Used BibTeX style is unsrt
\bibliography{ref}

\begin{thebibliography}{74}
\expandafter\ifx\csname natexlab\endcsname\relax\def\natexlab#1{#1}\fi

\bibitem[{{Allende Prieto} {et~al.}(2008){Allende Prieto}, {Majewski},
  {Schiavon}, {Cunha}, {Frinchaboy}, {Holtzman}, {Johnston}, {Shetrone},
  {Skrutskie}, {Smith}, \& {Wilson}}]{alendeprieto08}
{Allende Prieto}, C., {Majewski}, S.~R., {Schiavon}, R., {et~al.} 2008,
  Astronomische Nachrichten, 329, 1018

\bibitem[{Amarsi {et~al.}(2016)Amarsi, Lind, Asplund, Barklem, \&
  Collet}]{amarsi_non-lte_2016-1}
Amarsi, A.~M., Lind, K., Asplund, M., Barklem, P.~S., \& Collet, R. 2016,
  Monthly Notices of the Royal Astronomical Society, 463, 1518

\bibitem[{{Baines} {et~al.}(2018){Baines}, {Armstrong}, {Schmitt}, {Zavala},
  {Benson}, {Hutter}, {Tycner}, \& {van Belle}}]{baines18}
{Baines}, E.~K., {Armstrong}, J.~T., {Schmitt}, H.~R., {et~al.} 2018, \aj, 155,
  30

\bibitem[{{Baines} {et~al.}(2008){Baines}, {McAlister}, {ten Brummelaar},
  {Turner}, {Sturmann}, {Sturmann}, {Goldfinger}, \& {Ridgway}}]{Baines08}
{Baines}, E.~K., {McAlister}, H.~A., {ten Brummelaar}, T.~A., {et~al.} 2008,
  \apj, 680, 728

\bibitem[{{Bazot} {et~al.}(2018){Bazot}, {Creevey}, {Christensen-Dalsgaard}, \&
  {Mel{\'e}ndez}}]{Bazot18}
{Bazot}, M., {Creevey}, O., {Christensen-Dalsgaard}, J., \& {Mel{\'e}ndez}, J.
  2018, \aap, 619, A172

\bibitem[{{Bazot} {et~al.}(2011){Bazot}, {Ireland}, {Huber}, {Bedding},
  {Broomhall}, {Campante}, {Carfantan}, {Chaplin}, {Elsworth}, {Mel{\'e}ndez},
  {Petit}, {Th{\'e}ado}, {Van Grootel}, {Arentoft}, {Asplund}, {Castro},
  {Christensen-Dalsgaard}, {Do Nascimento}, {Dintrans}, {Dumusque}, {Kjeldsen},
  {McAlister}, {Metcalfe}, {Monteiro}, {Santos}, {Sousa}, {Sturmann},
  {Sturmann}, {ten Brummelaar}, {Turner}, \& {Vauclair}}]{Bazot11}
{Bazot}, M., {Ireland}, M.~J., {Huber}, D., {et~al.} 2011, \aap, 526, L4

\bibitem[{{Belokurov} {et~al.}(2020){Belokurov}, {Penoyre}, {Oh}, {Iorio},
  {Hodgkin}, {Evans}, {Everall}, {Koposov}, {Tout}, {Izzard}, {Clarke}, \&
  {Brown}}]{belokurov20}
{Belokurov}, V., {Penoyre}, Z., {Oh}, S., {et~al.} 2020, \mnras, 496, 1922

\bibitem[{{Bessell}(2000)}]{bessell00}
{Bessell}, M.~S. 2000, \pasp, 112, 961

\bibitem[{{Bovy}(2015)}]{Bovy15}
{Bovy}, J. 2015, \apjs, 216, 29

\bibitem[{{Boyajian} {et~al.}(2012{\natexlab{a}}){Boyajian}, {McAlister}, {van
  Belle}, {Gies}, {ten Brummelaar}, {von Braun}, {Farrington}, {Goldfinger},
  {O'Brien}, {Parks}, {Richardson}, {Ridgway}, {Schaefer}, {Sturmann},
  {Sturmann}, {Touhami}, {Turner}, \& {White}}]{Boyajian12a}
{Boyajian}, T.~S., {McAlister}, H.~A., {van Belle}, G., {et~al.}
  2012{\natexlab{a}}, \apj, 746, 101

\bibitem[{{Boyajian} {et~al.}(2014){Boyajian}, {van Belle}, \& {von
  Braun}}]{boyajian14}
{Boyajian}, T.~S., {van Belle}, G., \& {von Braun}, K. 2014, \aj, 147, 47

\bibitem[{{Boyajian} {et~al.}(2013){Boyajian}, {von Braun}, {van Belle},
  {Farrington}, {Schaefer}, {Jones}, {White}, {McAlister}, {ten Brummelaar},
  {Ridgway}, {Gies}, {Sturmann}, {Sturmann}, {Turner}, {Goldfinger}, \&
  {Vargas}}]{Boyajian13}
{Boyajian}, T.~S., {von Braun}, K., {van Belle}, G., {et~al.} 2013, \apj, 771,
  40

\bibitem[{{Boyajian} {et~al.}(2012{\natexlab{b}}){Boyajian}, {von Braun}, {van
  Belle}, {McAlister}, {ten Brummelaar}, {Kane}, {Muirhead}, {Jones}, {White},
  {Schaefer}, {Ciardi}, {Henry}, {L{\'o}pez-Morales}, {Ridgway}, {Gies}, {Jao},
  {Rojas-Ayala}, {Parks}, {Sturmann}, {Sturmann}, {Turner}, {Farrington},
  {Goldfinger}, \& {Berger}}]{Boyajian12b}
{Boyajian}, T.~S., {von Braun}, K., {van Belle}, G., {et~al.}
  2012{\natexlab{b}}, \apj, 757, 112

\bibitem[{{Brown} {et~al.}(1991){Brown}, {Gilliland}, {Noyes}, \&
  {Ramsey}}]{brown91}
{Brown}, T.~M., {Gilliland}, R.~L., {Noyes}, R.~W., \& {Ramsey}, L.~W. 1991,
  \apj, 368, 599

\bibitem[{{Casagrande} {et~al.}(2014){Casagrande}, {Portinari}, {Glass},
  {Laney}, {Silva Aguirre}, {Datson}, {Andersen}, {Nordstr{\"o}m}, {Holmberg},
  {Flynn}, \& {Asplund}}]{Casagrande14}
{Casagrande}, L., {Portinari}, L., {Glass}, I.~S., {et~al.} 2014, \mnras, 439,
  2060

\bibitem[{{Casagrande} \& {VandenBerg}(2014)}]{Casagrande_VandenBerg14}
{Casagrande}, L. \& {VandenBerg}, D.~A. 2014, \mnras, 444, 392

\bibitem[{{Casagrande} \& {VandenBerg}(2018)}]{Casagrande_VandenBerg18a}
{Casagrande}, L. \& {VandenBerg}, D.~A. 2018, \mnras, 475, 5023

\bibitem[{{Claret}(2000)}]{claret00}
{Claret}, A. 2000, \aap, 363, 1081

\bibitem[{{Claret} \& {Bloemen}(2011)}]{claret11}
{Claret}, A. \& {Bloemen}, S. 2011, \aap, 529, A75

\bibitem[{{de Jong} {et~al.}(2012){de Jong}, {Bellido-Tirado}, {Chiappini},
  {Depagne}, {Haynes}, {Johl}, {Schnurr}, {Schwope}, {Walcher}, {Dionies},
  {Haynes}, {Kelz}, {Kitaura}, {Lamer}, {Minchev}, {M{\"u}ller}, {Nuza},
  {Olaya}, {Piffl}, {Popow}, {Steinmetz}, {Ural}, {Williams}, {Winkler},
  {Wisotzki}, {Ansorge}, {Banerji}, {Gonzalez Solares}, {Irwin}, {Kennicutt},
  {King}, {McMahon}, {Koposov}, {Parry}, {Sun}, {Walton}, {Finger}, {Iwert},
  {Krumpe}, {Lizon}, {Vincenzo}, {Amans}, {Bonifacio}, {Cohen}, {Francois},
  {Jagourel}, {Mignot}, {Royer}, {Sartoretti}, {Bender}, {Grupp}, {Hess},
  {Lang-Bardl}, {Muschielok}, {B{\"o}hringer}, {Boller}, {Bongiorno}, {Brusa},
  {Dwelly}, {Merloni}, {Nandra}, {Salvato}, {Pragt}, {Navarro}, {Gerlofsma},
  {Roelfsema}, {Dalton}, {Middleton}, {Tosh}, {Boeche}, {Caffau}, {Christlieb},
  {Grebel}, {Hansen}, {Koch}, {Ludwig}, {Quirrenbach}, {Sbordone}, {Seifert},
  {Thimm}, {Trifonov}, {Helmi}, {Trager}, {Feltzing}, {Korn}, \&
  {Boland}}]{dejong12}
{de Jong}, R.~S., {Bellido-Tirado}, O., {Chiappini}, C., {et~al.} 2012, in
  Society of Photo-Optical Instrumentation Engineers (SPIE) Conference Series,
  Vol. 8446, Ground-based and Airborne Instrumentation for Astronomy IV, ed.
  I.~S. {McLean}, S.~K. {Ramsay}, \& H.~{Takami}, 84460T

\bibitem[{{De Silva} {et~al.}(2015){De Silva}, {Freeman}, {Bland-Hawthorn},
  {Martell}, {de Boer}, {Asplund}, {Keller}, {Sharma}, {Zucker}, {Zwitter},
  {Anguiano}, {Bacigalupo}, {Bayliss}, {Beavis}, {Bergemann}, {Campbell},
  {Cannon}, {Carollo}, {Casagrande}, {Casey}, {Da Costa}, {D'Orazi}, {Dotter},
  {Duong}, {Heger}, {Ireland}, {Kafle}, {Kos}, {Lattanzio}, {Lewis}, {Lin},
  {Lind}, {Munari}, {Nataf}, {O'Toole}, {Parker}, {Reid}, {Schlesinger},
  {Sheinis}, {Simpson}, {Stello}, {Ting}, {Traven}, {Watson}, {Wittenmyer},
  {Yong}, \& {{\v{Z}}erjal}}]{desilva15}
{De Silva}, G.~M., {Freeman}, K.~C., {Bland-Hawthorn}, J., {et~al.} 2015,
  \mnras, 449, 2604

\bibitem[{{Demarque} {et~al.}(2008){Demarque}, {Guenther}, {Li}, {Mazumdar}, \&
  {Straka}}]{Demarque08}
{Demarque}, P., {Guenther}, D.~B., {Li}, L.~H., {Mazumdar}, A., \& {Straka},
  C.~W. 2008, \apss, 316, 31

\bibitem[{{Derekas} {et~al.}(2011){Derekas}, {Kiss}, {Borkovits}, {Huber},
  {Lehmann}, {Southworth}, {Bedding}, {Balam}, {Hartmann}, {Hrudkova},
  {Ireland}, {Kov{\'a}cs}, {Mez{\H o}}, {Mo{\'o}r}, {Niemczura}, {Sarty},
  {Szab{\'o}}, {Szab{\'o}}, {Telting}, {Tkachenko}, {Uytterhoeven}, {Benk{\H
  o}}, {Bryson}, {Maestro}, {Simon}, {Stello}, {Schaefer}, {Aerts}, {ten
  Brummelaar}, {De Cat}, {McAlister}, {Maceroni}, {M{\'e}rand}, {Still},
  {Sturmann}, {Sturmann}, {Turner}, {Tuthill}, {Christensen-Dalsgaard},
  {Gilliland}, {Kjeldsen}, {Quintana}, {Tenenbaum}, \& {Twicken}}]{Derekas11}
{Derekas}, A., {Kiss}, L.~L., {Borkovits}, T., {et~al.} 2011, Science, 332, 216

\bibitem[{{Dotter} {et~al.}(2008){Dotter}, {Chaboyer}, {Jevremovi{\'c}},
  {Kostov}, {Baron}, \& {Ferguson}}]{Dotter2008}
{Dotter}, A., {Chaboyer}, B., {Jevremovi{\'c}}, D., {et~al.} 2008, \apjs, 178,
  89

\bibitem[{{ESA}(1997)}]{hipparcos}
{ESA}, ed. 1997, ESA Special Publication, Vol. 1200, {ESA, 1997, The HIPPARCOS
  and TYCHO catalogues}

\bibitem[{{Evans} {et~al.}(2018){Evans}, {Riello}, {De Angeli}, {Carrasco},
  {Montegriffo}, {Fabricius}, {Jordi}, {Palaversa}, {Diener}, {Busso},
  {Cacciari}, {van Leeuwen}, {Burgess}, {Davidson}, {Harrison}, {Hodgkin},
  {Pancino}, {Richards}, {Altavilla}, {Balaguer-N{\'u}{\~n}ez}, {Barstow},
  {Bellazzini}, {Brown}, {Castellani}, {Cocozza}, {De Luise}, {Delgado},
  {Ducourant}, {Galleti}, {Gilmore}, {Giuffrida}, {Holl}, {Kewley}, {Koposov},
  {Marinoni}, {Marrese}, {Osborne}, {Piersimoni}, {Portell}, {Pulone},
  {Ragaini}, {Sanna}, {Terrett}, {Walton}, {Wevers}, \&
  {Wyrzykowski}}]{evans18}
{Evans}, D.~W., {Riello}, M., {De Angeli}, F., {et~al.} 2018, \aap, 616, A4

\bibitem[{{Gaia Collaboration} {et~al.}(2016){Gaia Collaboration}, {Prusti},
  {de Bruijne}, {Brown}, {Vallenari}, {Babusiaux}, {Bailer-Jones}, {Bastian},
  {Biermann}, {Evans}, \& et~al.}]{gaia16}
{Gaia Collaboration}, {Prusti}, T., {de Bruijne}, J.~H.~J., {et~al.} 2016,
  \aap, 595, A1

\bibitem[{{Gilmore} {et~al.}(2012){Gilmore}, {Randich}, {Asplund}, {Binney},
  {Bonifacio}, {Drew}, {Feltzing}, {Ferguson}, {Jeffries}, {Micela}, \&
  et~al.}]{Gilmore12}
{Gilmore}, G., {Randich}, S., {Asplund}, M., {et~al.} 2012, The Messenger, 147,
  25

\bibitem[{{Green} {et~al.}(2015){Green}, {Schlafly}, {Finkbeiner}, {Rix},
  {Martin}, {Burgett}, {Draper}, {Flewelling}, {Hodapp}, {Kaiser}, {Kudritzki},
  {Magnier}, {Metcalfe}, {Price}, {Tonry}, \& {Wainscoat}}]{green15}
{Green}, G.~M., {Schlafly}, E.~F., {Finkbeiner}, D.~P., {et~al.} 2015, \apj,
  810, 25

\bibitem[{Gustafsson {et~al.}(2008)Gustafsson, Edvardsson, Eriksson,
  J{\o}rgensen, Nordlund, \& Plez}]{gustafsson_grid_2008}
Gustafsson, B., Edvardsson, B., Eriksson, K., {et~al.} 2008, Astronomy and
  Astrophysics, 486, 951

\bibitem[{{Guzik} {et~al.}(2016){Guzik}, {Houdek}, {Chaplin}, {Smalley},
  {Kurtz}, {Gilliland}, {Mullally}, {Rowe}, {Bryson}, {Still}, {Antoci},
  {Appourchaux}, {Basu}, {Bedding}, {Benomar}, {Garcia}, {Huber}, {Kjeldsen},
  {Latham}, {Metcalfe}, {P{\'a}pics}, {White}, {Aerts}, {Ballot}, {Boyajian},
  {Briquet}, {Bruntt}, {Buchhave}, {Campante}, {Catanzaro},
  {Christensen-Dalsgaard}, {Davies}, {Do{\u{g}}an}, {Dragomir}, {Doyle},
  {Elsworth}, {Frasca}, {Gaulme}, {Gruberbauer}, {Handberg}, {Hekker},
  {Karoff}, {Lehmann}, {Mathias}, {Mathur}, {Miglio}, {Molenda-{\.Z}akowicz},
  {Mosser}, {Murphy}, {R{\'e}gulo}, {Ripepi}, {Salabert}, {Sousa}, {Stello}, \&
  {Uytterhoeven}}]{Guzik16}
{Guzik}, J.~A., {Houdek}, G., {Chaplin}, W.~J., {et~al.} 2016, \apj, 831, 17

\bibitem[{{Heiter} {et~al.}(2015){Heiter}, {Jofr{\'e}}, {Gustafsson}, {Korn},
  {Soubiran}, \& {Th{\'e}venin}}]{heiter15}
{Heiter}, U., {Jofr{\'e}}, P., {Gustafsson}, B., {et~al.} 2015, \aap, 582, A49

\bibitem[{{Hj{\o}rringgaard} {et~al.}(2017){Hj{\o}rringgaard}, {Silva Aguirre},
  {White}, {Huber}, {Pope}, {Casagrande}, {Justesen}, \&
  {Christensen-Dalsgaard}}]{hjoerringgaard17}
{Hj{\o}rringgaard}, J.~G., {Silva Aguirre}, V., {White}, T.~R., {et~al.} 2017,
  \mnras, 464, 3713

\bibitem[{{H{\o}g} {et~al.}(2000){H{\o}g}, {Fabricius}, {Makarov}, {Urban},
  {Corbin}, {Wycoff}, {Bastian}, {Schwekendiek}, \& {Wicenec}}]{Hoeg2000}
{H{\o}g}, E., {Fabricius}, C., {Makarov}, V.~V., {et~al.} 2000, \aap, 355, L27

\bibitem[{{Huber} {et~al.}(2012){Huber}, {Ireland}, {Bedding}, {Brand{\~a}o},
  {Piau}, {Maestro}, {White}, {Bruntt}, {Casagrande}, {Molenda-{\.Z}akowicz},
  {Silva Aguirre}, {Sousa}, {Barclay}, {Burke}, {Chaplin},
  {Christensen-Dalsgaard}, {Cunha}, {De Ridder}, {Farrington}, {Frasca},
  {Garc{\'{\i}}a}, {Gilliland}, {Goldfinger}, {Hekker}, {Kawaler}, {Kjeldsen},
  {McAlister}, {Metcalfe}, {Miglio}, {Monteiro}, {Pinsonneault}, {Schaefer},
  {Stello}, {Stumpe}, {Sturmann}, {Sturmann}, {ten Brummelaar}, {Thompson},
  {Turner}, \& {Uytterhoeven}}]{huber12}
{Huber}, D., {Ireland}, M.~J., {Bedding}, T.~R., {et~al.} 2012, \apj, 760, 32

\bibitem[{{Ireland} {et~al.}(2008){Ireland}, {M{\'e}rand}, {ten Brummelaar},
  {Tuthill}, {Schaefer}, {Turner}, {Sturmann}, {Sturmann}, \&
  {McAlister}}]{Ireland2008}
{Ireland}, M.~J., {M{\'e}rand}, A., {ten Brummelaar}, T.~A., {et~al.} 2008, in
  \procspie, Vol. 7013, Optical and Infrared Interferometry, 701324

\bibitem[{{Jofr{\'e}} {et~al.}(2015){Jofr{\'e}}, {Heiter}, {Soubiran},
  {Blanco-Cuaresma}, {Masseron}, {Nordlander}, {Chemin}, {Worley}, {Van Eck},
  \& {Hourihane}}]{jofre15}
{Jofr{\'e}}, P., {Heiter}, U., {Soubiran}, C., {et~al.} 2015, \aap, 582, A81

\bibitem[{{Jofr{\'e}} {et~al.}(2014){Jofr{\'e}}, {Heiter}, {Soubiran},
  {Blanco-Cuaresma}, {Worley}, {Pancino}, {Cantat-Gaudin}, {Magrini},
  {Bergemann}, {Gonz{\'a}lez Hern{\'a}ndez}, {Hill}, {Lardo}, {de Laverny},
  {Lind}, {Masseron}, {Montes}, {Mucciarelli}, {Nordlander}, {Recio Blanco},
  {Sobeck}, {Sordo}, {Sousa}, {Tabernero}, {Vallenari}, \& {Van Eck}}]{Jofre14}
{Jofr{\'e}}, P., {Heiter}, U., {Soubiran}, C., {et~al.} 2014, \aap, 564, A133

\bibitem[{{Joyce} \& {Chaboyer}(2018)}]{Joyce18}
{Joyce}, M. \& {Chaboyer}, B. 2018, \apj, 864, 99

\bibitem[{{Kallinger} {et~al.}(2010){Kallinger}, {Weiss}, {Barban}, {Baudin},
  {Cameron}, {Carrier}, {De Ridder}, {Goupil}, {Gruberbauer}, {Hatzes},
  {Hekker}, {Samadi}, \& {Deleuil}}]{kallinger10}
{Kallinger}, T., {Weiss}, W.~W., {Barban}, C., {et~al.} 2010, \aap, 509, A77

\bibitem[{{Karovicova} {et~al.}(2020){Karovicova}, {White}, {Nordlander},
  {Casagrande}, {Ireland}, {Huber}, \& {Jofr{\'e}}}]{karovicova20}
{Karovicova}, I., {White}, T.~R., {Nordlander}, T., {et~al.} 2020, \aap, 640,
  A25

\bibitem[{{Karovicova} {et~al.}(2018){Karovicova}, {White}, {Nordlander},
  {Lind}, {Casagrande}, {Ireland}, {Huber}, {Creevey}, {Mourard}, {Schaefer},
  {Gilmore}, {Chiavassa}, {Wittkowski}, {Jofr{\'e}}, {Heiter}, {Th{\'e}venin},
  \& {Asplund}}]{karovicova18}
{Karovicova}, I., {White}, T.~R., {Nordlander}, T., {et~al.} 2018, \mnras, 475,
  L81

\bibitem[{{Kervella} {et~al.}(2019){Kervella}, {Arenou}, {Mignard}, \&
  {Th{\'e}venin}}]{kervella19}
{Kervella}, P., {Arenou}, F., {Mignard}, F., \& {Th{\'e}venin}, F. 2019, \aap,
  623, A72

\bibitem[{{Kjeldsen} \& {Bedding}(1995)}]{kjeldsen95}
{Kjeldsen}, H. \& {Bedding}, T.~R. 1995, \aap, 293, 87

\bibitem[{{Klinglesmith} \& {Sobieski}(1970)}]{klinglesmith70}
{Klinglesmith}, D.~A. \& {Sobieski}, S. 1970, \aj, 75, 175

\bibitem[{{Ligi} {et~al.}(2016){Ligi}, {Creevey}, {Mourard}, {Crida},
  {Lagrange}, {Nardetto}, {Perraut}, {Schultheis}, {Tallon-Bosc}, \& {ten
  Brummelaar}}]{Ligi16}
{Ligi}, R., {Creevey}, O., {Mourard}, D., {et~al.} 2016, \aap, 586, A94

\bibitem[{{Lin} {et~al.}(2018){Lin}, {Dotter}, {Ting}, \& {Asplund}}]{Lin18}
{Lin}, J., {Dotter}, A., {Ting}, Y.-S., \& {Asplund}, M. 2018, \mnras, 477,
  2966

\bibitem[{{Lund} {et~al.}(2017){Lund}, {Silva Aguirre}, {Davies}, {Chaplin},
  {Christensen-Dalsgaard}, {Houdek}, {White}, {Bedding}, {Ball}, {Huber},
  {Antia}, {Lebreton}, {Latham}, {Handberg}, {Verma}, {Basu}, {Casagrande},
  {Justesen}, {Kjeldsen}, \& {Mosumgaard}}]{Lund17}
{Lund}, M.~N., {Silva Aguirre}, V., {Davies}, G.~R., {et~al.} 2017, \apj, 835,
  172

\bibitem[{{Maestro} {et~al.}(2013){Maestro}, {Che}, {Huber}, {Ireland},
  {Monnier}, {White}, {Kok}, {Robertson}, {Schaefer}, {ten Brummelaar}, \&
  {Tuthill}}]{Maestro13}
{Maestro}, V., {Che}, X., {Huber}, D., {et~al.} 2013, \mnras, 434, 1321

\bibitem[{{Magic} {et~al.}(2015){Magic}, {Chiavassa}, {Collet}, \&
  {Asplund}}]{magic15}
{Magic}, Z., {Chiavassa}, A., {Collet}, R., \& {Asplund}, M. 2015, \aap, 573,
  A90

\bibitem[{{Magic} {et~al.}(2013){Magic}, {Collet}, {Asplund}, {Trampedach},
  {Hayek}, {Chiavassa}, {Stein}, \& {Nordlund}}]{magic13}
{Magic}, Z., {Collet}, R., {Asplund}, M., {et~al.} 2013, \aap, 557, A26

\bibitem[{{Mason} {et~al.}(2001){Mason}, {Wycoff}, {Hartkopf}, {Douglass}, \&
  {Worley}}]{mason01}
{Mason}, B.~D., {Wycoff}, G.~L., {Hartkopf}, W.~I., {Douglass}, G.~G., \&
  {Worley}, C.~E. 2001, \aj, 122, 3466

\bibitem[{{Minchev} {et~al.}(2018){Minchev}, {Anders}, {Recio-Blanco},
  {Chiappini}, {de Laverny}, {Queiroz}, {Steinmetz}, {Adibekyan}, {Carrillo},
  {Cescutti}, {Guiglion}, {Hayden}, {de Jong}, {Kordopatis}, {Majewski},
  {Martig}, \& {Santiago}}]{minchev18}
{Minchev}, I., {Anders}, F., {Recio-Blanco}, A., {et~al.} 2018, \mnras, 481,
  1645

\bibitem[{{Moultaka} {et~al.}(2004){Moultaka}, {Ilovaisky}, {Prugniel}, \&
  {Soubiran}}]{Moultaka2004}
{Moultaka}, J., {Ilovaisky}, S.~A., {Prugniel}, P., \& {Soubiran}, C. 2004,
  \pasp, 116, 693

\bibitem[{{Neckel} \& {Labs}(1994)}]{NeckelLabs94}
{Neckel}, H. \& {Labs}, D. 1994, \solphys, 153, 91

\bibitem[{{O'Donnell}(1994)}]{odonnell94}
{O'Donnell}, J.~E. 1994, \apj, 422, 158

\bibitem[{{Pereira} {et~al.}(2013){Pereira}, {Asplund}, {Collet}, {Thaler},
  {Trampedach}, \& {Leenaarts}}]{pereira13}
{Pereira}, T.~M.~D., {Asplund}, M., {Collet}, R., {et~al.} 2013, \aap, 554,
  A118

\bibitem[{Piskunov \& Valenti(2017)}]{piskunov_spectroscopy_2017}
Piskunov, N. \& Valenti, J.~A. 2017, Astronomy and Astrophysics, 597, A16

\bibitem[{{Quirrenbach} {et~al.}(1996){Quirrenbach}, {Mozurkewich}, {Buscher},
  {Hummel}, \& {Armstrong}}]{quirrenbach96}
{Quirrenbach}, A., {Mozurkewich}, D., {Buscher}, D.~F., {Hummel}, C.~A., \&
  {Armstrong}, J.~T. 1996, \aap, 312, 160

\bibitem[{{Rabus} {et~al.}(2019){Rabus}, {Lachaume}, {Jord{\'a}n}, {Brahm},
  {Boyajian}, {von Braun}, {Espinoza}, {Berger}, {Le Bouquin}, \&
  {Absil}}]{rabus19}
{Rabus}, M., {Lachaume}, R., {Jord{\'a}n}, A., {et~al.} 2019, \mnras, 484, 2674

\bibitem[{{Raghavan} {et~al.}(2006){Raghavan}, {Henry}, {Mason}, {Subasavage},
  {Jao}, {Beaulieu}, \& {Hambly}}]{raghavan06}
{Raghavan}, D., {Henry}, T.~J., {Mason}, B.~D., {et~al.} 2006, \apj, 646, 523

\bibitem[{{Rains} {et~al.}(2020){Rains}, {Ireland}, {White}, {Casagrande}, \&
  {Karovicova}}]{rains20}
{Rains}, A.~D., {Ireland}, M.~J., {White}, T.~R., {Casagrande}, L., \&
  {Karovicova}, I. 2020, \mnras, 493, 2377

\bibitem[{{Rains} {et~al.}(2021){Rains}, {{\v{Z}}erjal}, {Ireland},
  {Nordlander}, {Bessell}, {Casagrande}, {Onken}, {Joyce}, {Kammerer}, \&
  {Abbot}}]{rains21}
{Rains}, A.~D., {{\v{Z}}erjal}, M., {Ireland}, M.~J., {et~al.} 2021, \mnras,
  504, 5788

\bibitem[{{Randich} {et~al.}(2013){Randich}, {Gilmore}, \& {Gaia-ESO
  Consortium}}]{Randich13}
{Randich}, S., {Gilmore}, G., \& {Gaia-ESO Consortium}. 2013, The Messenger,
  154, 47

\bibitem[{{Schwarzschild}(1906)}]{Schwarzschild1906}
{Schwarzschild}, K. 1906, Nachr. K{\"o}nigl. Ges. Wiss. G{\"o}ttingen,
  Math.-Phys. Kl., 195, 41

\bibitem[{{Silva Aguirre} {et~al.}(2017){Silva Aguirre}, {Lund}, {Antia},
  {Ball}, {Basu}, {Christensen-Dalsgaard}, {Lebreton}, {Reese}, {Verma},
  {Casagrande}, {Justesen}, {Mosumgaard}, {Chaplin}, {Bedding}, {Davies},
  {Handberg}, {Houdek}, {Huber}, {Kjeldsen}, {Latham}, {White}, {Coelho},
  {Miglio}, \& {Rendle}}]{SilvaAguirre17}
{Silva Aguirre}, V., {Lund}, M.~N., {Antia}, H.~M., {et~al.} 2017, \apj, 835,
  173

\bibitem[{{Skrutskie} {et~al.}(2006){Skrutskie}, {Cutri}, {Stiening},
  {Weinberg}, {Schneider}, {Carpenter}, {Beichman}, {Capps}, {Chester},
  {Elias}, {Huchra}, {Liebert}, {Lonsdale}, {Monet}, {Price}, {Seitzer},
  {Jarrett}, {Kirkpatrick}, {Gizis}, {Howard}, {Evans}, {Fowler}, {Fullmer},
  {Hurt}, {Light}, {Kopan}, {Marsh}, {McCallon}, {Tam}, {Van Dyk}, \&
  {Wheelock}}]{Skrutskie2006}
{Skrutskie}, M.~F., {Cutri}, R.~M., {Stiening}, R., {et~al.} 2006, \aj, 131,
  1163

\bibitem[{{Tayar} {et~al.}(2020){Tayar}, {Claytor}, {Huber}, \& {van
  Saders}}]{tayar20}
{Tayar}, J., {Claytor}, Z.~R., {Huber}, D., \& {van Saders}, J. 2020, arXiv
  e-prints, arXiv:2012.07957

\bibitem[{{ten Brummelaar} {et~al.}(2005){ten Brummelaar}, {McAlister},
  {Ridgway}, {Bagnuolo}, {Turner}, {Sturmann}, {Sturmann}, {Berger}, {Ogden},
  {Cadman}, {Hartkopf}, {Hopper}, \& {Shure}}]{tenBrummelaar05}
{ten Brummelaar}, T.~A., {McAlister}, H.~A., {Ridgway}, S.~T., {et~al.} 2005,
  \apj, 628, 453

\bibitem[{{Ulrich}(1986)}]{ulrich86}
{Ulrich}, R.~K. 1986, \apjl, 306, L37

\bibitem[{{von Braun} {et~al.}(2014){von Braun}, {Boyajian}, {van Belle},
  {Kane}, {Jones}, {Farrington}, {Schaefer}, {Vargas}, {Scott}, {ten
  Brummelaar}, {Kephart}, {Gies}, {Ciardi}, {L{\'o}pez-Morales}, {Mazingue},
  {McAlister}, {Ridgway}, {Goldfinger}, {Turner}, \& {Sturmann}}]{vonbraun14}
{von Braun}, K., {Boyajian}, T.~S., {van Belle}, G.~T., {et~al.} 2014, \mnras,
  438, 2413

\bibitem[{{White} {et~al.}(2011){White}, {Bedding}, {Stello},
  {Christensen-Dalsgaard}, {Huber}, \& {Kjeldsen}}]{White11}
{White}, T.~R., {Bedding}, T.~R., {Stello}, D., {et~al.} 2011, \apj, 743, 161

\bibitem[{{White} {et~al.}(2013){White}, {Huber}, {Maestro}, {Bedding},
  {Ireland}, {Baron}, {Boyajian}, {Che}, {Monnier}, {Pope}, {Roettenbacher},
  {Stello}, {Tuthill}, {Farrington}, {Goldfinger}, {McAlister}, {Schaefer},
  {Sturmann}, {Sturmann}, {ten Brummelaar}, \& {Turner}}]{white13}
{White}, T.~R., {Huber}, D., {Maestro}, V., {et~al.} 2013, \mnras, 433, 1262

\bibitem[{{White} {et~al.}(2018){White}, {Huber}, {Mann}, {Casagrande},
  {Grunblatt}, {Justesen}, {Silva Aguirre}, {Bedding}, {Ireland}, {Schaefer},
  \& {Tuthill}}]{white18}
{White}, T.~R., {Huber}, D., {Mann}, A.~W., {et~al.} 2018, \mnras, 477, 4403

\end{thebibliography}

 \begin{appendix}
\section{Limb-darkening coefficients in 38 channels.}
\begin{table}[h!]\small
\caption{Limb-darkening coefficients in 38 channels. We show one table for the star HD\,131156 as an example.
Limb-darkening coefficients for the rest of the stars are available at CDS in tables A.1.--A.9.}   \label{table:A1}
 \centering                          
\begin{tabular}{llllll}      
\hline\hline  
\multicolumn{5}{l}{HD\,131156}  \\
&\multicolumn{4}{c}{four-term limb-darkening$^a$} \\   
chan.&   wavelength  &    $a_1$ & $a_2$ & $a_3$ & $a_4$          \\%$\theta_\mathrm{LD}$ (mas) \\ 
\hline    
1.   &   0.630     &    0.551  &      0.050    &    0.337  &     -0.178    \\
2.   &   0.635     &    0.561  &      0.026    &    0.353  &     -0.183    \\
3.   &   0.639     &    0.565  &      0.016    &    0.366  &     -0.193    \\
4.   &   0.644     &    0.575  &     -0.016    &    0.395  &     -0.202    \\
5.   &   0.649     &    0.571  &     -0.014    &    0.403  &     -0.213    \\
6.   &   0.654     &    0.588  &      0.024    &    0.299  &     -0.174    \\
7.   &   0.659     &    0.585  &      0.026    &    0.303  &     -0.180    \\
8.   &   0.664     &    0.583  &     -0.022    &    0.379  &     -0.199    \\
9.   &   0.670     &    0.582  &     -0.022    &    0.365  &     -0.187    \\
10.  &   0.675     &    0.583  &     -0.020    &    0.352  &     -0.179    \\
11.  &   0.680     &    0.588  &     -0.037    &    0.363  &     -0.183    \\
12.  &   0.686     &    0.592  &     -0.055    &    0.384  &     -0.193    \\
13.  &   0.692     &    0.589  &     -0.041    &    0.360  &     -0.184    \\
14.  &   0.698     &    0.596  &     -0.067    &    0.382  &     -0.190    \\
15.  &   0.704     &    0.604  &     -0.095    &    0.411  &     -0.203    \\
16.  &   0.710     &    0.599  &     -0.073    &    0.374  &     -0.185    \\
17.  &   0.716     &    0.605  &     -0.100    &    0.397  &     -0.193    \\
18.  &   0.722     &    0.599  &     -0.081    &    0.367  &     -0.178    \\
19.  &   0.729     &    0.602  &     -0.098    &    0.388  &     -0.190    \\
20.  &   0.736     &    0.611  &     -0.119    &    0.404  &     -0.196    \\
21.  &   0.743     &    0.611  &     -0.126    &    0.400  &     -0.192    \\
22.  &   0.750     &    0.612  &     -0.126    &    0.399  &     -0.193    \\
23.  &   0.756     &    0.613  &     -0.125    &    0.388  &     -0.188    \\
24.  &   0.763     &    0.620  &     -0.153    &    0.415  &     -0.200    \\
25.  &   0.771     &    0.617  &     -0.139    &    0.389  &     -0.187    \\
26.  &   0.778     &    0.622  &     -0.155    &    0.398  &     -0.189    \\
27.  &   0.786     &    0.620  &     -0.148    &    0.394  &     -0.191    \\
28.  &   0.794     &    0.618  &     -0.150    &    0.392  &     -0.189    \\
29.  &   0.802     &    0.618  &     -0.153    &    0.388  &     -0.185    \\
30.  &   0.810     &    0.616  &     -0.147    &    0.381  &     -0.183    \\
31.  &   0.818     &    0.618  &     -0.162    &    0.386  &     -0.183    \\
32.  &   0.827     &    0.616  &     -0.157    &    0.381  &     -0.182    \\
33.  &   0.835     &    0.617  &     -0.168    &    0.389  &     -0.186    \\
34.  &   0.844     &    0.614  &     -0.158    &    0.374  &     -0.181    \\
35.  &   0.853     &    0.663  &     -0.319    &    0.507  &     -0.219    \\
36.  &   0.863     &    0.646  &     -0.262    &    0.453  &     -0.203    \\
37.  &   0.872     &    0.627  &     -0.208    &    0.395  &     -0.181    \\
38.  &   0.881     &    0.627  &     -0.207    &    0.396  &     -0.184    \\
\hline                                 
\end{tabular}%\tablefoot{$^{(a)}$ Limb-darkening coefficients derived from the grid of \citet{magic15}; see text for details.}
\flushleft $^{a}$ Limb-darkening coefficients derived from the grid of \citet{magic15}; see text for details.
\end{table}

\end{appendix}

\end{document}